\documentclass[aps,pra,twocolumn,floatfix,showpacs]{revtex4-1}
\usepackage{graphicx,bm}
\usepackage{times}
\usepackage{amsmath,amssymb}

\begin{document}
\title{Quantification of memory effect of steady-state current from interaction-induced transport in quantum systems}

\author{Chen-Yen Lai}
\author{Chih-Chun Chien}
\affiliation{School of Natural Sciences, University of California, Merced, California 95343, USA.}

\date{\today}

\begin{abstract}
  Dynamics of a system in general depends on its initial state and how the system is driven, but in many-body systems the memory is usually averaged out during evolution.
  Here, interacting quantum systems without external relaxations are shown to retain long-time memory effects in steady states.
  To identify memory effects, we first show quasi-steady state currents form in finite, isolated Bose and Fermi Hubbard models driven by interaction imbalance and they become steady-state currents in the thermodynamic limit.
  By comparing the steady state currents from different initial states or ramping rates of the imbalance, long-time memory effects can be quantified.
  While the memory effects of initial states are more ubiquitous, the memory effects of switching protocols are mostly visible in interaction-induced transport in lattices.
  Our simulations suggest the systems enter a regime governed by a generalized Fick's law and memory effects lead to initial-state dependent diffusion coefficients.
  We also identify conditions for enhancing memory effects and discuss possible experimental implications.
\end{abstract}


\maketitle

\section{Introduction}
Although dynamic processes generally depend on the initial conditions and evolution protocols, many-body systems tend to average out, in the long-time limit, memory of initial information.
Take metals as an example, the fast relaxation time quickly brings the electrons to a new equilibrium or steady state after a perturbation~\cite{Schoenlein:1987dg,Roberti:1995du,Bigot:1995he}.
Therefore, systems exhibiting long-time or persistent memory effects, for instance magnetic hysteresis~\cite{bertotti1998hysteresis}, shape-memory materials~\cite{Jagla:2015vd}, and memory-effect elements and circuitries~\cite{Pershin:2011ie,Marani:2015we}, are considered interesting.
Moreover, artificial spin ice driven by a cyclic magnetic field shows memory effects of reproducible microstates~\cite{Gilbert:2015vx}.
In a partial symmetry breaking Hamiltonian, the symmetry memory and symmetry gap show quantum memory effects of initial states below a critical value~\cite{Zhao:2017up}.
Other examples of memory effects in quantum systems include ferroelectric semiconductor~\cite{Folcia:1987bs,Aliyev:2003jn} and magnetic materials~\cite{Marti:2014fl,Sun:2003iz}.

Moreover, the concept of memory effects is related to the existence of  fundamental limits in many-body quantum systems on the relaxation of current~\cite{Maldacena:2016gp} and development of thermal equilibrium~\cite{Polkovnikov:2011iu}.
For example, systems exhibiting many-body localization~\cite{Pasienski:2010bd,Meldgin:2016cu,Kondov:2015ce,Fischer:2015ve} can fail to thermalize, show ergodicity breaking~\cite{Schreiber:2015jt}, and retain local information about the initial condition at long times~\cite{Smith:2015uo}.
The out-of-time-order correlation functions~\cite{Tsuji:2017hm,Dora:2017go,Haehl:2017ui} provide another useful measure to diagnose the sensitivity of time-evolving quantities on the initial condition and offer a tool to investigate fast scrambling of information~\cite{Yao:2016tb}.
Cold-atom systems with their broadly tunable parameters are suitable for elucidating key mechanisms behind quantum transport~\cite{Chien:2015kc}.
Memory effects, in the form of hysteresis, have been explored in cold-atom experiments of atomic superfluids~\cite{Eckel:2014gf} and theoretical proposal of spin-orbit coupled Fermi gases~\cite{Metcalf:2016dm}.
Energy dissipation has been included by vortex generation or external reservoirs in those hysteresis studies.
In addition, memory effects of open quantum systems coupled classical and quantum environments have also been intensely studied~\cite{PhysRevA.95.052126,PhysRevA.92.012315}.
Here we will investigate memory effects in isolated quantum systems without explicit relaxations.

In the absence of interactions, dynamical variables of quantum systems, such as the mass current, are usually found not to exhibit memory effects in steady states~\cite{Cornean:2013kz,Chien:2013ef, Chien:2013en, Chien:2014uf}.
To identify long-time memory effects, the density of a tunable bound state may provide clues in the steady state~\cite{Cornean:2013kz}.
A flat band coming from localized states after a continuous transformation of the underlying lattice geometry also reveals memory effects of the transformation rate in the steady-state density distribution~\cite{Lai:2016hr}.
Moreover, memory effects of entanglement and correlation can arise from topological edge states as boundary conditions~\cite{He:2016fc,Metcalf:2017vya,DaWang:2015hr} or as parameters~\cite{Chung:2013ce} are changed.
In addition, quench dynamics of integrable systems can exhibit memory effects, where a steady state emerges and depends strongly on the initial condition~\cite{Rigol:2006df,Rigol:2007bm,Chung:2012bl}.
In contrast, interacting quantum systems can exhibit memory effects directly in the steady-state current as illustrated in Luttinger liquids~\cite{Perfetto:2010ih}, where currents driven by different quench procedures reach different steady-state values.

Here we quantify long-time memory effects in the steady-state current of isolated interacting quantum systems, including both fermionic and bosonic gases.
Since transport usually singles out a particular direction, we consider one-dimensional lattice systems where particle hopping due to tunneling slows down the motion.
To focus on intrinsic effects, the current can be induced, for example, internally by switching on a spatial interaction imbalance, so there is no need for external source and sink.
In a finite system, the boundary reflects the wavefunction and leads to revival time which linearly scales with the system size.
However, a quasi-steady state current (QSSC) emerges before the revival time and allows us to quantify memory effects by comparing the steady-state values from different driving protocols.
As the system size $L$ scales toward the thermodynamic limit, the revival time scales toward infinity ($L\rightarrow\infty$ and then $t\rightarrow\infty$).
The quasi-steady state currents (QSSCs) then become genuine steady state currents and memory effects revealed by the different values of currents persist in the long-time limit.
In isolated systems discussed here, the transient regime before the system enters the quasi-steady state does not scale with the system size.
Therefore, the steady state in the thermodynamic limit is similar to the quasi-steady state of a large but finite system and memory effects can be identified by comparing the  steady-state currents.

\begin{figure}[bh]
  \begin{center}
    \includegraphics[width=0.48\textwidth]{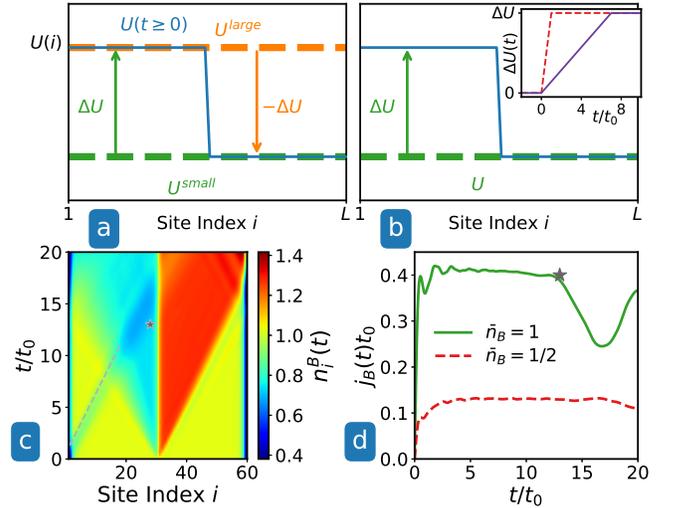}
    \caption{
      Interaction induced transport for probing (a) memory effects of initial states and (b) memory effects of switching protocols.
      (a) Two systems initially with different uniform interactions $U^{\text{small}}$ (green dashed) and $U^{\text{large}}$ (orange dashed) are quenched to the same interaction profile (blue solid).
      (b) Systems with the same initial interaction (green dashed) experience the same interaction imbalance (blue solid) switched on at different time scales.
      The inset shows the linear switching protocols with different ramping times ($t_r=t_0$, $7t_0$).
      (c)-(d) Interaction-induced transport in the BHM.
      (c) Particle density contour plot with $L\!=\!60$ sites, $\overline{n}_B\!=\!1$, $U\!=\!J$, and $\Delta U\!=\!J$ on the left half of the lattice.
      The dashed line in (c) indicates the propagation of the low-density region from the left edge and the star marks the time the density distortion affects the QSSC, as one can see the QSSC decreases at the star mark in (d).
      The red (light blue) regime on the right (left) half of the system indicates formation of density plateaus.
      The star marks in (c) and (d) indicate the same time ($t\!=\!13t_0$).
      (d) Current flowing through the interface of interaction imbalance with different fillings and $U\!=\!\Delta U\!=\!J$.
    }
    \label{fig:1}
  \end{center}
\end{figure}

Two types of memory effects will be analyzed here, one from different initial states and the other from different protocols of inducing transport. For instance, an interaction imbalance or potential imbalance can be applied.
To distinguish the two memory effects, we use different ways to turn on the imbalance.
For the first type of memory effects, we consider systems with different initial ground states but quenched to the same final interaction or potential profile as illustrated in Fig.~\ref{fig:1}(a).
For the second type, the same imbalance is switched on linearly with different time scales ($t_r$) as shown in Fig.~\ref{fig:1}(b).
Here we present the results of interaction-induced transport in lattices and discuss continuous models and potential-driven transport.
While memory effects of initial states are visible in all of those settings, memory effects of switching protocols are mostly visible only in interaction-induced transport in lattices.

The paper is organized as follows.
Section~\ref{sec:transport} introduces the models and simulation methods, and then we present the results from interaction-induced transport.
Section~\ref{sec:mem} discusses memory effects of initial states and memory effects of switching protocols.
The continuous model is discussed in Sec.~\ref{sec:gp} and serves as a comparison.
Section~\ref{sec:ficks} presents a connection between Fick's law and our studies and the diffusion coefficients extracted from the relation.
Section~\ref{sec:vinduce} is devoted to potential-induced transport and its memory effects.
In Section~\ref{sec:exp} we discuss experimental implications of cold atoms, possible implementations, and how memory effects of steady-state currents may be measured.
A conclusion is give in Sec.~\ref{sec:conclude}.

\begin{figure}[t]
  \begin{center}
    \includegraphics[width=0.48\textwidth]{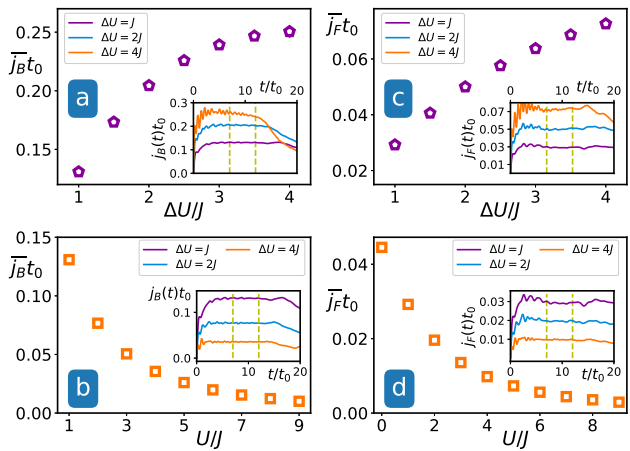}
    \caption{
      QSSCs in the BHM with $\overline{n}_B\!=\!1/2$ and FHM with $\overline{n}_F\!=\!1/6$.
      The insets shows the evolution of mass current where the plateaus indicate the QSSC, and dashed lines indicate the time window where the averages are taken..
      (a) BHM and (c) FHM with the same uniform initial interaction $U\!=\!J$ and different interaction imbalances.
      (b) BHM and (d) FHM with different uniform initial interactions and the same interaction imbalance $\Delta U\!=\!J$.
      A stronger initial interaction suppresses the QSSC value.
      The error bars are the standard deviation from the time average and they are within the symbol size if not shown.
    }
    \label{fig:T1}
  \end{center}
\end{figure}

\begin{figure}[t]
  \begin{center}
    \includegraphics[width=0.48\textwidth]{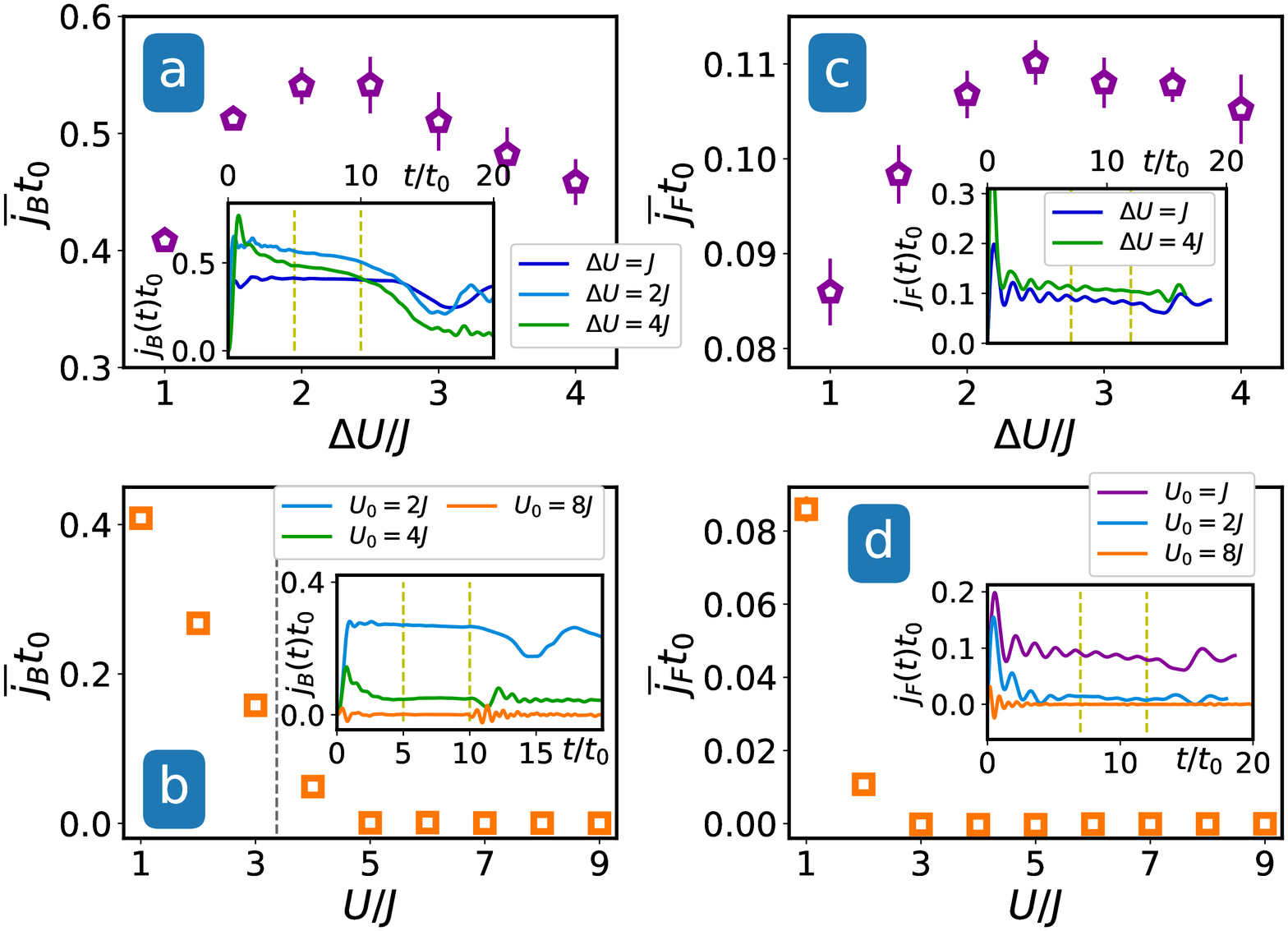}
    \caption{
      Interaction-induced transport by a sudden interaction imbalance of the BHM  with $\overline{n}_B\!=\!1$ in (a)-(b) and the FHM with $\overline{n}_F\!=\!1/2$ in (c)-(d).
      The system size is $L\!=\!60$ sites.
      Under a constant initial interaction $U\!=\!J$, the QSSCs as a function of the interaction imbalance are show in (a) for the BHM and (c) for the FHM.
      By quenching the systems with the same $\Delta U\!=\!J$ on the left half, the QSSCs decrease with the initial interaction as shown in (b) for BHM and (d) for FHM.
      The dashed line in (b) indicates the critical interaction of 1D superfluid-Mott insulator transition.
      The dashed lines in the insets indicate the time window where the average is taken.
    }
    \label{fig:T2}
  \end{center}
\end{figure}

\section{Interaction-induced transport}\label{sec:transport}
A mass current can be induced by imposing an interaction imbalance as illustrated in Fig.~\ref{fig:1}(a) and (b).
We consider two lattices models:
The Hamiltonian of the Bose Hubbard model (BHM) is given by
\begin{equation}\label{eq:bhm}
	\mathcal{H}_{\text{BHM}}=-J\sum_{\langle i,j\rangle}(b^\dagger_ib_j+h.c.)+\sum_i\frac{U(i,t)}{2}n_i(n_i-1).
\end{equation}
Here $b^\dagger_i$ $(b_i)$ is the boson creation (annihilation) operator on lattice site $i$, $n_i\!=\! b^\dagger_ib_i$ is the boson number operator on site $i$, and $\langle i,j\rangle$ represents nearest neighbors.
We set $\hbar\!=\!1$ and the time unit is $t_0\!=\!\hbar/J$.
The Fermi Hubbard model (FHM) has the Hamiltonian
\begin{equation}\label{eq:fhm}
	\mathcal{H}_{\text{FHM}}=-J\sum_{\langle i,j\rangle,\sigma}(c^\dagger_{i,\sigma}c_{j,\sigma}+h.c.)+\sum_iU(i,t)n_{i,\uparrow}n_{i,\downarrow}.
\end{equation}
The two components of fermions are usually two different hyperfine states of the same species of atoms and we use  $\sigma=\uparrow,\downarrow$ to label them.
Here $c^\dagger_{i,\sigma}$ $(c_{i,\sigma})$ is the fermion creation (annihilation) operator on lattice site $i$ with spin $\sigma$, $n_{i,\sigma}\!=\! c^\dagger_{i,\sigma}c_{i,\sigma}$ is the fermion number operator on site $i$ with spin $\sigma$.
The filling of the BHM is $\overline{n}_B\!=\!\sum_i n_i/L$.
We consider $\overline{n}_B$ up to $1$ for a  system with $L\!=\!60$ sites and monitor the current at the interaction-imbalance interface, so  $j_B\!=\!2J\text{Im}\langle b^\dagger_{L/2+1}b_{L/2}\rangle$.
As for the FHM, we consider the spin-balanced case where $N_{\uparrow}\!=\!\sum_in_{i,\sigma}\!=\!N_{\downarrow}$ with filling $\overline{n}_F\!=\!N_{\sigma}/L$ up to $1/2$, and the current is the total current $j_F\!=\!2J\text{Im}\langle\sum_{\sigma} c^\dagger_{L/2+1,\sigma}c_{L/2,\sigma}\rangle$.

Initially, an uniform interaction is applied throughout the system and the ground state can be obtained by the density matrix renormalization group (DMRG)~\cite{White:1992ie,White:1993fb,Ostlund:1995fx,Schollwock:2005jv,McCulloch:2007gi,Schollwock:2011gl}.
Then, a spatial interaction imbalance on half of the system is imposed, as illustrated in Fig.~\ref{fig:1}.
A positive (negative) interaction imbalance  $\Delta U$ is applied to the left (right) half of the system so the final Hamiltonian always has a larger interaction energy on the left half of the system.
The dynamics can be simulated with the time-dependent DMRG~\cite{White:2004fd,Schollwock:2005jv,Lai:2008ez,Schollwock:2011gl,Vidal:2004jc} using the second order Suzuki-Trotter formula with a time step $\delta t=0.005t_0$.
During the simulation, the maximum bond dimension is kept up to $\chi\!=\!1000$ states and we are able to maintain the maximal truncation error below $10^{-7}$.
Most of the simulations are performed with system size $L\!=\!60$.
By comparing with larger $L$, we have ensured that $L$ is large enough for the same result to be qualitatively observed in the thermodynamic limit.

A typical example of the interaction-induced dynamics is shown in Fig.~\ref{fig:1}(c) and (d).
Right after the interaction imbalance is applied, the system passes through a transient regime ($t\!<\!4t_0$), and then the density on the left and right halves start to develop plateaus ($t\!=\!10t_0$) as show in Figs.~\ref{fig:1}(c) and~\ref{fig:ficks}(a).
The existence of the density plateaus is important for the quasi-steady state current (QSSC) to emerge.
As the plateaus develop, the density gradient around the interface of the interaction imbalance will be constant when the QSSC is observable.
This indicates the system may be describable by coarse-grained kinetic-equations~\cite{kamenev2011field,DiVentra:2010ks} connecting the density and current.
We emphasize the transport is induced by changing the coupling strength in the Hamiltonian only, and there is no external exchange of energy or particle because the system is isolated.

The results of interaction-induced transport are summarized in Figs.~\ref{fig:T1} and~\ref{fig:T2}.
QSSCs can be observed as the plateaus in the insets of Figs.~\ref{fig:T1} and~\ref{fig:T2}, and they will become genuine steady-states in the thermodynamic limit.
While the QSSC can be observed in interacting bosonic and fermionic systems, its existence in noninteracting systems is less trivial.
Homogeneous noninteracting bosons do not support QSSC~\cite{Chien:2012cv}, and we observe no QSSC if half of the bosonic system is noninteracting during the dynamics which will be discussed later in this section.
For fermions, QSSCs already exist in the absence of interactions~\cite{Chien:2013ef,DiVentra:2004bx}.
If $\overline{n}_F\! < \!1/2$ or $\overline{n}_B\! < \!1$, a larger interaction imbalance results in a larger QSSC under the same initial condition.
On the other hand, the QSSC value gets smaller but never reaches zero as the initial interaction is stronger while the interaction imbalance is fixed, as shown in Fig.~\ref{fig:T1}(b) and (d).

\subsection{Additional features at higher filling}
At or above $\overline{n}_F\!=\!1/2$ or $\overline{n}_B\!=\!1$, strong interactions lead to additional phenomena.
For example, negative differential conductance (NDC), where the current decreases as the driving force increases, has been discussed in fermions~\cite{Chien:2013ef} and observed in bosons~\cite{Labouvie:2015dx}.
In isolated systems, energy conservation can lead to dynamically insulating phases~\cite{Chien:2013ef,Lai:2016kh}.
Moreover, the Mott-insulating phase at $\overline{n}_B\!=\!1$ or $\overline{n}_F\!=\!1/2$ with moderate initial interactions can be destroyed by a strong potential bias~\cite{Oka:2003hr,Oka:2005dk,HeidrichMeisner:2010fr} or interaction imbalance.

\begin{figure*}[t]
	\begin{center}
		\includegraphics[width=0.78\textwidth]{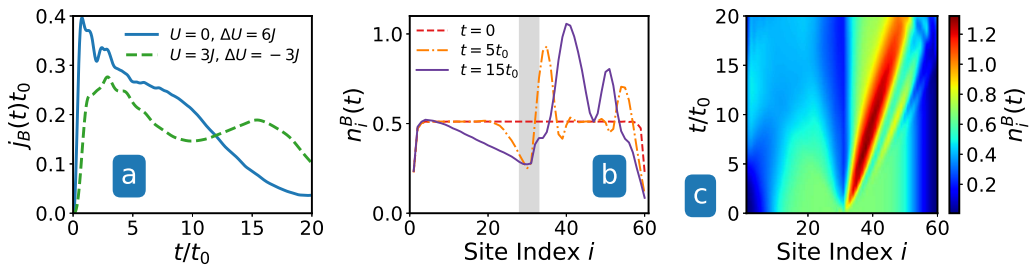}
		\caption{
			No QSSC can be observed if part of a bosonic system is noninteracting.
			(a) Currents through the middle of a lattice with $L\!=\!60$ sites and $\overline{n}_B\!=\!1/2$ versus time for tow cases: Initially noninteracting bosons experiencing an interaction imbalance (solid blue) and initially interacting bosons with part of the system quenched to noninteracting bosons (dashed green).
			(b) Particle density profiles at selected times for a system with an initial interaction $U\!=\!3J$ and an imbalance $\Delta U\!=\!-3J$ applied on the right half.
			There is no steady structures in the central (shaded) region when compared to Fig.~\ref{fig:ficks}(a).
			(c) Evolution of the particle density for $U\!=\!0$ and $\Delta U\!=\!6J$ on the left half.
		}
		\label{fig:NoSteady}
	\end{center}
\end{figure*}

\subsubsection{Bose Hubbard Model at unity filling}
The main results of BHM at unity filling are presented in Fig.~\ref{fig:T2}(a)-(b), where the symbols show the time-average of QSSCs and the statistical standard deviation within a selected time window is indicated by the error bar.
As a larger interaction imbalance shrinks the duration of QSSC due to the boundary effect, we did not observe the negative differential conductivity (NDC) when filling is lower than one within the parameter range we searched.
If the filling is unity, the Mott-superfluid transition occurs at the critical interaction $U_c\!\approx\!3.37J$ in one dimension~\cite{Kuhner:2000tg,Zakrzewski:2008gs}.

In Fig.~\ref{fig:T2}(a), the system is initially a superfluid with $U\!=\!J$ and the QSSCs do not increase monotonically with $\Delta U$, which is different from the behavior at low fillings.
The QSSCs increase with the interaction imbalance when $\Delta U$ is small, but the dependence changes once the interaction imbalance is above $2.5J$, beyond which the response (current) starts to decrease as the driving (interaction imbalance) increases.
Similar phenomena are also discovered in theory~\cite{Conwell:2008kf,Lai:2016kh} and cold-atom experiments~\cite{Labouvie:2015dx,Conwell:2008kf}.
For filling larger than unity, the systems are expected to suffer NDC as well when the interaction imbalance increases.
The origin of NDC is because the interaction imbalance causes an energy difference between the two sides of the system.
Since the isolated system respects energy conservation, changes in the interaction energy have to be compensated for by the kinetic energy.
As the interaction difference gets larger, it is harder for the two sides to exchange particles in an energy-conserved fashion.
Eventually, if the energy difference between the two parts of the system greatly exceeds the band width, a dynamically non-conducting state emerges because it is not possible to exchange particles without violating energy conservation in an isolated system.
The boundary effect limits our simulations with strong interaction imbalance as shown in the inset of Fig.~\ref{fig:T2}, where the error bars become larger due to the distortion of wavefunction from the edge.

If the interaction energy is within the range of bandwidth, the system does not become dynamically insulating if the initial state is in the superfluid regime. However, the QSSC becomes smaller as the initial interaction increases with the same  quenched interaction imbalance.
A finite QSSC is still observable for a system with initial interaction $U\!=\!4J\!>\!U_c$ in the Mott-insulator regime and experiencing an interaction imbalance $\Delta U\!=\!J$ on the left as shown in Fig.~\ref{fig:T2}(b).
This indicates a dynamic breakdown of the Mott insulating state by an interaction imbalance.
The reason for the breakdown is that at the interface of the interaction imbalance it is preferable to have a doublon-holon pair (with a hole on the stronger-interaction side and two bosons in one site on the weaker-interaction side) and lower the local interaction energy.
Creating this pair of excitations requires more energy as the initial interaction increases, and eventually a fixed $\Delta U$ is no longer sufficient to produce the pair.
Thus, the QSSC vanishes in the strong interaction regime as shown in Fig.~\ref{fig:T2}(b).
The destruction of the Mott insulating state has also been addressed by applying a strong potential imbalance in a Mott insulator~\cite{HeidrichMeisner:2010fr,Oka:2010ku} and the destruction can be understood by a many-body Schwinger-Landau-Zener mechanism~\cite{Oka:2003hr,Oka:2005dk}.

\subsubsection{Fermi Hubbard Model at half filling}
For the FHM with filling smaller than half, the transport behavior is qualitatively the same as the BHM discussed in Sec.~\ref{sec:transport}, except QSSCs exist in fermions even in the absence of interactions~\cite{Chien:2013ef}.
The 1D FHM has a charge gap for any non-zero interaction $U\!>\!0$ at half filling in the thermodynamic limit according to the exact solution~\cite{Lieb:2003hl}.
Thus, both the NDC and dynamically insulating state are expected to happen in the FHM at half filling, and indeed they are shown in Fig.~\ref{fig:T2}(c) and (d).
Similar to the BHM, with small initial interactions like $U\!=\!J$ or $2J$, it is possible to break the Mott insulating state by a strong interaction imbalance through the mechanism of a pair of doublon and holon, similar to the BHM case.
When the initial interaction or the energy imbalance greatly exceeds the bandwidth, energy conservation prevents exchange of particles with large energy difference and the system will remain an insulator.

\subsection{Absence of QSSC in noninteracting bosons}
For bosonic systems, we found that QSSCs can only exist when the interactions on both halves are finite.
To check the absence of QSSC if part of the system is noninteracting, we consider two cases here: One starts from a noninteracting system and then the left half is quenched to a finite interaction.
The second is initially an uniform, interacting system and the interaction on the right half is turned off.
The results are shown in Fig.~\ref{fig:NoSteady}(a), and there is no QSSC in both cases.
The reason that noninteracting Bose gases cannot support QSSC is due to the infinite compressibility which allows bosons to pile up without limit if there is no interaction energy.
According to the kinetic-equation approach~\cite{kamenev2011field, DiVentra:2010ks} shown in Eq.~\eqref{eq:diffusion} below, the QSSCs depend on a constant density gradient at the interface.
If half of the system is noninteracting, the bosons will keep stacking up and never reach a plateau as shown in Fig.~\ref{fig:NoSteady}(b).
This can also be observed by comparing Figs.~\ref{fig:NoSteady}(c) and~\ref{fig:1}(c).

\section{Memory effects}\label{sec:mem}
By comparing the QSSCs induced by different protocols, we can quantify memory effects in the steady state.
Previous studies of interaction-induced transport~\cite{Chien:2013ef} using a mean-field approximation failed to see memory effects because multi-particle correlations are ignored, but here we found two main causes of memory effects: the initial state and the protocol for switching-on the interaction imbalance.
\begin{figure}[t]
  \begin{center}
    \includegraphics[width=0.48\textwidth]{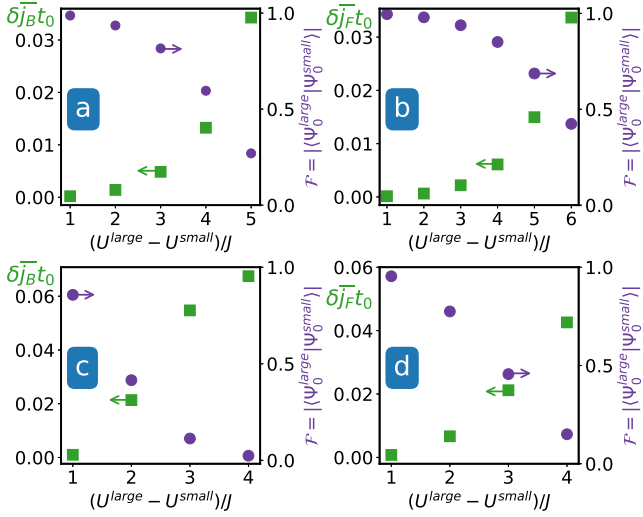}
    \caption{
      Memory effects of initial states for the BHM with (a) $\overline{n}_{B}\!=\!1/2$ and (c) $\overline{n}_{B}\!=\!1$ and for the FHM with (b) $\overline{n}_{F}\!=\!1/6$, and (d) $\overline{n}_{F}\!=\!1/2$.
      The square symbols correspond to the difference of steady-state currents from two initial interactions $U^{\text{large}}=6J$ and $U^{\text{small}}$, which can be inferred from the $(U^{\text{large}}-U^{\text{small}})/J$ values, quenched to the same final interaction profile.
      The circles show the fidelity $\mathcal{F}$ between the two initial ground states.
    }
    \label{fig:MemInit}
  \end{center}
\end{figure}

\begin{figure}[t]
  \begin{center}
    \includegraphics[width=0.48\textwidth]{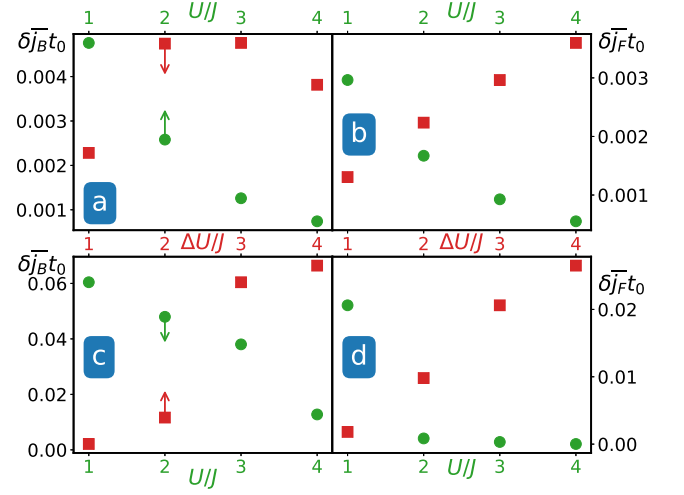}
    \caption{
      Memory effects of switching protocols for BHM with (a)  $\overline{n}_{B}\!=\!1/2$ and (c) $\overline{n}_{B}\!=\!1$ and for the FHM with (b) $\overline{n}_{F}\!=\!1/6$, and (d) $\overline{n}_{F}\!=\!1/2$.
      Memory effects of switching protocols are probed with different finite ramping times $t_r\!=\!t_0$ and $7t_0$ for varying interaction imbalance but fixed initial interaction $U\!=\!J$ (red squares) and for different initial interaction but the same interaction imbalance $\Delta U\!=\!3J$ (green circles).
      A larger difference of the averaged steady-state currents indicates stronger memory effects.
      Starting with $U\!=\!J$, memory effects become stronger as the interaction imbalance increases.
      Memory effects are suppressed when the initial interaction is strong.
      For $t_r\!=\!t_0$ ($7t_0$), the averages are taken between $5t_0$-$10t_0$ ($10t_0$-$15t_0$).
    }
    \label{fig:MemTime}
  \end{center}
\end{figure}

\subsection{Memory effects of initial states}\label{sec:meminit}
To analyze the memory effects of initial states, we introduce different interaction imbalances so that different initial configurations are suddenly changed to the same final configuration, and the system evolves accordingly.
As illustrated in Fig.~\ref{fig:1}(a), one can consider two initial configurations, one with $U(i)\!=\!U^{\text{small}}$ and one with $U(i)\!=\!U^{\text{large}},\forall i$.
The difference between the two initial ground states can be quantified by the fidelity $\mathcal{F}\!=\!\vert\langle\Psi_0^{\text{large}}\vert\Psi_0^{\text{small}}\rangle\vert$.
Then, sudden quenches at $t\!=\!0$ are applied to the two systems so that $\Delta U\!=\!U^{\text{large}}-U^{\text{small}}\!>\!0$ is applied to the left half of the first system, and $-\Delta U$ is applied to the right half of the second system.
After the quenches, both systems have the same interaction profile with $U^{\text{large}}$ ($U^{\text{small}}$) on the left (right) half.
Comparing the QSSCs of the two systems then reveals the memory effect from initial states.

The difference of steady-state values of the currents and the fidelity of the initial states are shown in Fig.~\ref{fig:MemInit}(a) for bosons with $\overline{n}_B\!=\!1/2$ and (b) for fermions with $\overline{n}_F\!=\!1/6$.
Here, the current is averaged over a time period within the steady-state regime: $\overline{j_{B(F)}}\!=\!\sum_{t=t_i}^{t_f}\Delta tj_{B(F)}(t)/(t_f-t_i)$ with $\Delta t\!=\!0.02t_0$, and the difference of average currents between two initial states are $\delta\overline{j_{B(F)}}\!=\!\lvert\overline{j_{B(F)}^{\text{large}}}-\overline{j_{B(F)}^{\text{small}}}\rvert$.
As one can see, even though the final Hamiltonians are identical, the steady-state currents do not necessarily agree.
A larger difference of the averaged current indicates stronger memory effects.
In the same plot, the fidelity between the two initial states shows that the steady-state memory effects are stronger as the two initial states have less overlap.
Interestingly, there is no qualitative difference between the bosonic and fermionic results.
The correlations from interactions, which are missing in mean-field theory~\cite{Chien:2013ef}, are important for correctly describing memory effects.

The BHM with $\overline{n}_B\!=\!1$ is a Mott insulator for the system initially with $U^{\text{large}}\!=\!6J$ and the one initially with $U^{\text{small}}\!=\!5J$, so the QSSC induced by $\Delta U\!=\!J$ is zero and no memory effects can be identified in Fig.~\ref{fig:MemInit}(c).
When using a smaller $U^{\text{small}}$ (a stronger $\Delta U$) as the initial condition, the fidelity between the two initial wavefunctions decreases, and memory effects start to emerge.
After comparing different fillings with the same parameter settings, we found bosons at unity filling exhibit stronger memory effects than at low fillings.
Similar phenomena are also observed in the FHM as shown in Fig.~\ref{fig:MemInit}(d).

\subsection{Memory effects of switching protocols}\label{sec:memtime}
The second kind of memory effects arises from different switching protocols, and here we focus on its dependence on the ramping time when the interaction imbalance is switched on linearly in time.
Taking two systems with the same initial and final configurations but with two different ramping time scales as illustrated in Fig.~\ref{fig:1}(b) and its inset, the memory effects from different ramping time scales can then be identified from the different QSSCs.
The results are show in Fig.~\ref{fig:MemTime}, where we chose two different time scales $t_r\!=\!t_0$ and $7t_0$ and compare the steady-state values.
The figure shows difference of average currents between two different time scales, $\delta\overline{j_{B(F)}}\!=\!\lvert\overline{j_{B(F)}^{7t_0}}-\overline{j_{B(F)}^{t_0}}\rvert$.
We have checked other values of $t_r$ between $5t_0$ and $7t_0$, and there is no qualitative difference. We also found, in general, fermions exhibit stronger memory effects.
On the other hand, if the ramping of interaction imbalance becomes slower and eventually reach the adiabatic limit, i.e. $t_r\!\rightarrow\!\infty$, there will be no finite QSSCs as the system remains in the equilibrium state during time evolution.

Fig.~\ref{fig:MemTime} offers clues on enhancing the memory effects of switching protocols.
With the initial interaction fixed, a larger interaction imbalance tends to cause stronger memory effects regardless of the filling.
For a finite system, a stronger interaction imbalance can lead to stronger boundary effects and limit the time a QSSC can be maintained.
Thus, the memory effects of switching protocols are more visible in interaction-induced transport with a larger interaction imbalance and smaller initial interaction.

For the BHM at unity filling and the FHM at half filling, the general conclusion still holds as memory effects of steady-state currents are amplified when applying a stronger interaction imbalance with a weaker initial interaction.
However, both BHM and FHM will suffer the dynamically insulating state in the large interaction-imbalance regime, and there we cannot find QSSC or resolve memory effects in our simulations.
In the regimes where QSSC can be sustained, our results show that stronger memory effects can occur when the BHM is at unity filling and the FHM is at half filling as shown in Fig.~\ref{fig:MemTime}(c) and (d).

While memory effects of switching protocols are observable in interaction-induced transport in both BHM and FHM, they are more fragile in other settings partly because short-time correlations are usually averaged out.
The system enters the regime described by kinetic equations and is insensitive to its transient behavior~\cite{kamenev2011field,DiVentra:2010ks,2008entropy,altland2010condensed}.

\section{Continuum Model}\label{sec:gp}
\begin{figure}[t]
	\begin{center}
		\includegraphics[width=0.47\textwidth]{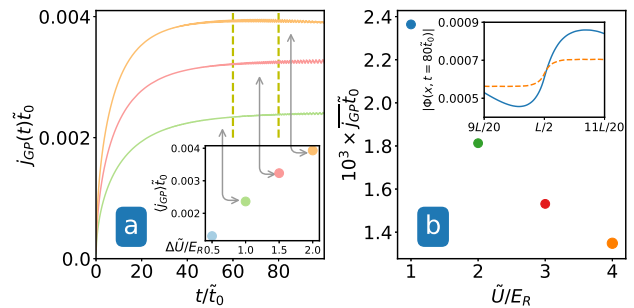}
		\caption{
			Dynamics of the GPE for $N_b\!=\!10$ bosons under (a) a constant initial interaction $\tilde{U}\!=\!E_R$ and (b) a constant interaction imbalance $\Delta \tilde{U}\!=\!E_R$.
			The results show decreasing QSSCs with increasing initial interaction or decreasing interaction imbalance.
			The statistical errors are within the symbol size and the dashed lines indicates the regime where the average is taken.
			The inset in (a) shows the averaged currents for different interaction imbalances.
			The inset in (b) shows the amplitudes of the condensate wavefunction near the center of the system (shaded regime) at time $t\!=\!80\tilde{t}_0$ for two different initial interactions $\tilde{U}\!=\!E_R$ (dashed) and $\tilde{U}\!=\!4E_R$ (solid).
		}
		\label{fig:GP}
	\end{center}
\end{figure}
To verify the ubiquity of memory effects, we also consider weakly interacting bosons in the continuum.
The zero-temperature Bose-Einstein condensate and its dynamics may be studied by the mean-field Gross-Pitaevskii equation (GPE)~\cite{Gross:1961bx,Pitaevsk:fe}.
It can be generalized to describe time-dependent systems~\cite{Vudragovic:2012jz,Pethick:2010gy,Stoof:2008ho} and has been previously implemented in modeling coherent transport~\cite{Rab:2008kp,Bradly:2012fc}.
In one dimension, the time-dependent GPE can be written as
\begin{equation}\label{eq:gpe}
  \left[-\frac{\hbar^2}{2m}\frac{d^2}{dx^2}+V_{ext}(x)+\tilde{U}(x,t)N_b|\Phi|^2\right]\Phi = i\hbar\frac{\partial}{\partial t}\Phi ,
\end{equation}
where $\Phi(r,t)$ is the condensate wave function, $m$ is the mass of the bosonic atom, and $N_b$ is the number of bosons.
The coupling constant $\tilde{U}\!=\! 4\pi\hbar^2a_s/m$ is determined by the two-body $s$-wave scattering length $a_s$.
In our simulation, the external potential $V_{ext}(x)$ corresponds to a box potential which confines the atoms.
Here we solve the GPE with algorithms involving real- and imaginary-time propagation based on a split-step Crank-Nicolson method~\cite{Muruganandam:2009dq,Vudragovic:2012jz} and follow Ref.~\cite{Cerimele:2000hs} to normalize the wavefunction with $\int dx |\Phi(x)|^2\!=\!1$.

\begin{figure}[h]
	\begin{center}
		\includegraphics[width=0.48\textwidth]{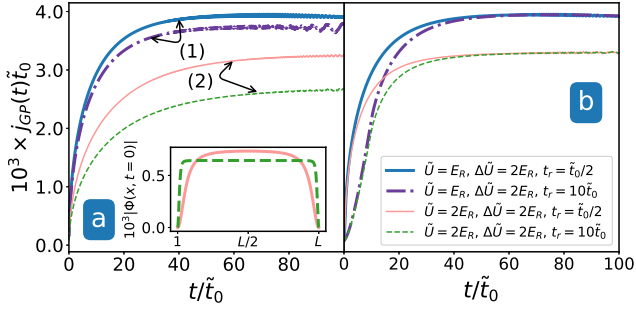}
		\caption{
			Memory effects of weakly interacting Bose gases.
			(a) Two examples with different initial conditions.
			Both cases reach the same final interaction configuration with $\tilde{U}^{\text{large}}$ on the left half and $\tilde{U}^{\text{small}}$ on the right half.
			In case (1), $\tilde{U}^{\text{large}}\!=\!3E_R$ (purple dashed dot) and $\tilde{U}^{\text{small}}\!=\!E_R$ (blue thick solid) are compared.
			In case (2), $\tilde{U}^{\text{large}}\!=\!1.5E_R$ (green dashed) and $\tilde{U}^{\text{small}}\!=\!0.5E_R$ (red thin solid) are compared.
			The two examples show the QSSCs depend on the initial conditions.
			The inset shows the initial profiles of the condensate wavefunctions for $\tilde{U}\!=\!0.5E_R$ (red solid) and $\tilde{U}\!=\!1.5E_R$ (green dashed), respectively.
			(b) The ramping of an interaction imbalance with finite time scales $t_r=\tilde{t}_0/2$, $10\tilde{t}_0$ for systems with $N_b\!=\!10$ bosons for the two cases $\tilde{U}\!=\!\Delta\tilde{U}/2\!=\!E_R$ (top two lines) and $\tilde{U}\!=\!\Delta\tilde{U}\!=\!2E_R$ (bottom two lines).
			The QSSC values from different ramping times cannot be distinguished in both cases.
		}
		\label{fig:GPmemory}
	\end{center}
\end{figure}

\subsection{Quasi-steady state current}
The initial state is the ground state of a system with size $l$ and an uniform interaction $\tilde{U}$.
The energy unit is the recoil energy $E_R\!=\!\pi^2\hbar^2/2ml^2$ and the time unit is $\tilde{t}_0\!=\!\hbar/E_R$.
At time $t\!=\!0$, the interaction imbalance is suddenly imposed with $\Delta\tilde{U}$ on the left half of the system, and we monitor the current flowing through the middle, which can be calculated by
\begin{equation}\label{eq:jgp}
  j_{GP}(t)=\frac{d}{dt}\int_{0}^{L/2}dxN_b|\Phi(x,t)|^2.
\end{equation}
From Fig.~\ref{fig:GP}(a), the current reaches a plateau indicating the existence of QSSC.
The duration of the QSSC scales as the system size increases, and the steady-state value allows us to quantify memory effects unambiguously.
We calculate the averaged current $\overline{j_{GP}}=\sum_{t=t_i}^{t_f}\Delta tj_{GP}(t)/(t_f-t_i)$ in a time window $(t_i,t_f)$ where the QSSC lasts.
Under the same initial interaction, the QSSC increases as a larger interaction imbalance is applied, as shown in the inset of Fig.~\ref{fig:GP}(a).
Macroscopically, the current is driven by the pressure difference between the two sides of the interaction imbalance.
On the other hand, the QSSC decreases as the initial interaction is stronger, as shown in Fig.~\ref{fig:GP}(b).
This is because a larger interaction presents a larger initial pressure and applying the same interaction imbalance triggers a smaller percentage change in the pressure.
While the transport properties of the GPE are similar to the BHM, the GPE does not exhibit the superfluid-Mott insulator transition.

\subsection{Memory effects of the continuum model}
The memory effects of initial states can be observed in the two examples of GPE shown in Fig.~\ref{fig:GPmemory}(a).
We notice that memory effects of initial states are more prominent in the dilute limit (with a small $N_b$) as the two initial interactions provide more distinct density profiles shown in the inset of Fig.~\ref{fig:GPmemory}(a).
If the density is dense (with a large $N_b$) and the two interactions are too strong in the beginning, the resulting initial states can be very similar and the memory effects would be difficult to resolve.
On the other hand, in Fig.~\ref{fig:GPmemory} (b) we show two examples which start with the same initial condition and reach the same final configuration but with different ramping time $t_r$.
In both examples, the steady-state values are the same and there is no observable memory effect.

For memory effects of switching schemes to survive, the short-time correlations should not be averaged out completely.
This requires the Green's function to depend sensitively on short-time changes.
However, this is unlikely the case in the continuum limit because the minimum difference of the intrinsic time scale is determined by $\Delta\tilde{t}_{\text{min}}\!\sim\!\hbar/\Delta E_{\text{max}}$, where $E_{\text{max}}$ may be identified as the width of the energy spectrum.
Since $E_{\text{max}}$ has no upper bound in the continuum model but is bounded by the bandwidth in lattice models, the Green's function always reduces to the ``short memory" approximation in the continuum model because one $t_0$ corresponds to many $\Delta\tilde{t}_{\text{min}}$ as the system evolves.
As a consequence, short-time correlations may survive and cause memory effects of switching protocols in lattice models, but they are mostly averaged out in the continuum.

\begin{figure}[b]
  \begin{center}
    \includegraphics[width=0.48\textwidth]{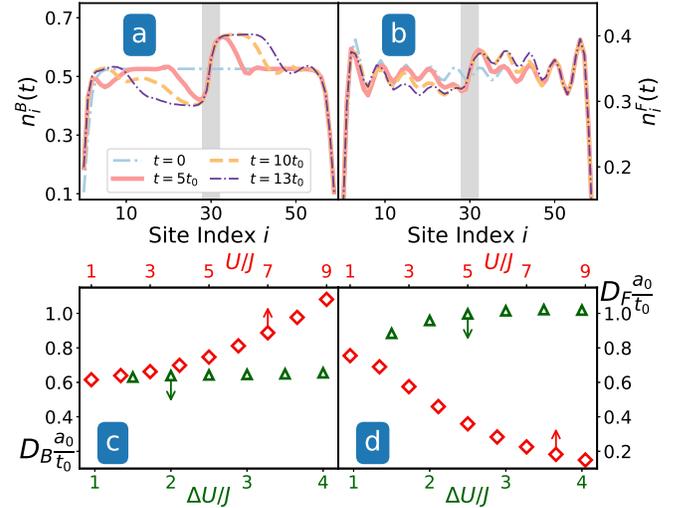}
    \caption{
     Evolution of density profiles and diffusion coefficients of (a) and (c) BHM with filling $\overline{n}_B\!=\!1/2$, and (b) and (d) FHM with $\overline{n}_F\!=\!1/6$ on a $60$-site lattice.
     A constant density gradient appears near the interface of interaction imbalance (shaded region) when the steady-state current lasts.
     (a) and (b) start with $U\!=\!J$ and undergo an interaction imbalance quench $\Delta U\!=\!J$ on the right half.
     (c) and (d) fix $\Delta U\!=\!J$ (or $U\!=\!J$) and vary $U$ (or $\Delta U$).
     Here $a_0$ is the lattice constant.
    }
    \label{fig:ficks}
  \end{center}
\end{figure}
\section{Fick's law and generalized diffusion coefficient}\label{sec:ficks}
In the absence of interactions, the transport is ballistic and one may use the Landauer theory~\cite{Landauer:cw,DiVentra:2010ks} to evaluate the current if the system is connected to two particle reservoirs.
On the other hand, in interacting systems the numerous energy levels may serve as a reservoir for the system itself~\cite{kamenev2011field}. After averaging out the short-term correlations, one may obtain a quantum kinetic equation for the density rather than the wavefunction~\cite{kamenev2011field,2008entropy}.
In one dimension, it has the form similar to the Fick's law:
\begin{equation}\label{eq:diffusion}
	j(x,t)=-D(x, t)\frac{\partial}{\partial x} n(x, t).
\end{equation}
However, the diffusion coefficient $D(x,t)$ can be a function of time and position and may be obtained from the space-time integral of the real-time Green's function $G(x-x^\prime;t-t^\prime)$.
For instance,
$D(x,t)=\frac{\int^{t}_{-\infty}w(t^\prime)dt^\prime\int G(x-x^\prime;t-t^\prime)dx}{w(t)\int G(x-x^\prime;t-t^\prime)dx}$,
where $w(t)$ describes how the driving is introduced.
The classical diffusion limit may be obtained by using a slow varying function in the interval $(-\infty,t)$ for $w(t)$ with additional assumptions of linearity, long-wavelength restriction characteristic of simple fluids, and the time interval $t-t^\prime$ being on the order of the mean free time.

At this point, the equation is still complicated, so it is common to use the ``short-memory" approximation.
Thus, only the very recent information is relevant to the present behavior.
This approximation simplifies Eq.~\eqref{eq:diffusion} to the classical diffusion equation, where $D(x,t)$ reduces to a constant and is identified as the diffusion coefficient.
Eq.~\eqref{eq:diffusion} is an approximation of the full quantum description where short-time correlations contribute to the coefficient $D(x,t)$.

The steady states observed in our simulations suggest the systems may be described by kinetic equations~\cite{kamenev2011field,DiVentra:2010ks,2008entropy,altland2010condensed}, where an effective description of the density rather than the wavefunction can be applied.
By analyzing the evolution of density profiles following a suddenly induced interaction imbalance, we identify a constant density gradient across the interface of interaction imbalance when the steady-state current lasts, as illustrated in the shaded regions in Fig.~\ref{fig:ficks}(a) and (b) for the BHM and FHM and the inset of Fig.~\ref{fig:GP}(b) for the continuum model.

The Fick's law relates the mass current and density gradient and implies diffusion-like behavior.
Assuming the isolated systems in the quasi-steady states follow the kinetic equations, we can estimate the diffusion coefficient defined by
\begin{equation}\label{eq:Diffusion}
  D_{B(F)}=\frac{\overline{j_{B(F)}}}{\overline{\nabla_i n_i^{B(F)}}}.
\end{equation}
The averages are taken in the same time window, the spatial gradient is taken along the lattice, and only the steady-state density gradient at the interaction-imbalance interface is considered.
In Fig.~\ref{fig:ficks}(c) and (d), we show the diffusion coefficients extracted from the BHM and FHM following a sudden quench of interaction imbalance.
Importantly, due to memory effects of initial states, the diffusion coefficients are sensitive to the initial conditions even if the final Hamiltonians are the same.
Therefore, in isolated interacting systems the diffusion coefficients can inherit long-time memory of initial states.

The diffusion coefficient exhibits interesting dependence on the spin-statistics.
With a fixed initial interaction, the diffusion coefficient increases with interaction imbalance in both spin statistics, although the variation in the BHM is very limited.
On the other hand, with a fixed interaction imbalance, the diffusion coefficient of the BHM (FHM) increases (decreases) monotonically as the initial interaction increases.
Our results, however, do not exclude possibilities that the transport behavior is more complicated than kinetic equations or their generalizations.

We remark that Eq.~\eqref{eq:Diffusion} may be more common than previously thought.
For instance, the spreading of a single-particle Gaussian wave packet has both time and initial state dependence.
The wavefunction of a single-particle Gaussian wave is packet~\cite{sakurai2011modern}
\begin{equation}\label{eq:wpf}
\psi(x,t)=\left[\frac{1}{\sigma[1+i(t/\tau)]\sqrt{2\pi}}\right]^{\frac{1}{2}}\exp\left\{-\frac{1}{4}\frac{(x-a)^2}{\sigma^2[1+i(t/\tau)]}\right\}.
\end{equation}
The initial condition has a normal distribution with mean $\langle x\rangle\!=\!a$ and variance $\langle(x-a)^2\rangle\!=\!\sigma^2$, and we define $\tau\!=\!2m\sigma^2/\hbar^2$.
The probability distribution is
$P(x,t)=\vert\psi(x,t)\vert^2=\frac{1}{\sigma(t)\sqrt{2\pi}}\exp\left\{-\frac{1}{2}[\frac{x-a}{\sigma(t)}]^2\right\}$,
where $\sigma(t)\!=\!\sigma\sqrt{1+(t/\tau)^2}$.
This wave packet has a fixed mean ($x\!=\!a$, $\forall t$) but a growing variance. The current density can be obtained from  Eq.~\eqref{eq:wpf} as
\begin{eqnarray}
j&=&\frac{\hbar}{2mi}\left[\psi^*\nabla\psi-(\nabla\psi^*)\psi\right]
=-\frac{\hbar(t/\tau)}{2m}\nabla P(x,t).
\end{eqnarray}
By using $\nabla P(x,t)=-\frac{x-a}{\sigma^2(t)}P(x,t)$ and comparing the result with Eq.~\eqref{eq:diffusion}, we conclude that
\begin{equation}
D(t)=\frac{\hbar^3}{4m^2\sigma^2}t.
\end{equation}
The diffusion coefficient obtained here is a time-dependent function instead of a constant.

\begin{figure*}[th]
  \begin{center}
    \includegraphics[width=0.78\textwidth]{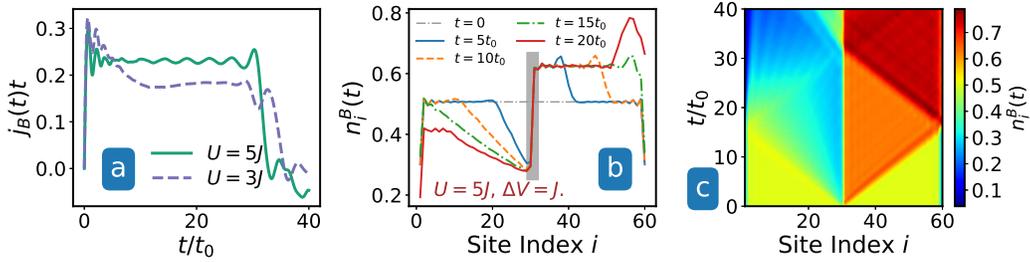}
    \caption{
      QSSC in potential-induced transport in the BHM with $\overline{n}_B\!=\!1/2$.
      The onsite potential energy on the left half of the system experiences a sudden change of $\Delta V\!=\!J$.
      (a) Current versus time for two systems with different initial interactions $U\!=3J$ (dashed) and $U\!=\!5J$ (solid).
      (b) Density profiles at different times and (c) density contour for a system with initial interaction $U\!=\!5J$.
    }
    \label{fig:RampVTransport}
  \end{center}
\end{figure*}

\begin{figure}[b]
  \begin{center}
    \includegraphics[width=0.48\textwidth]{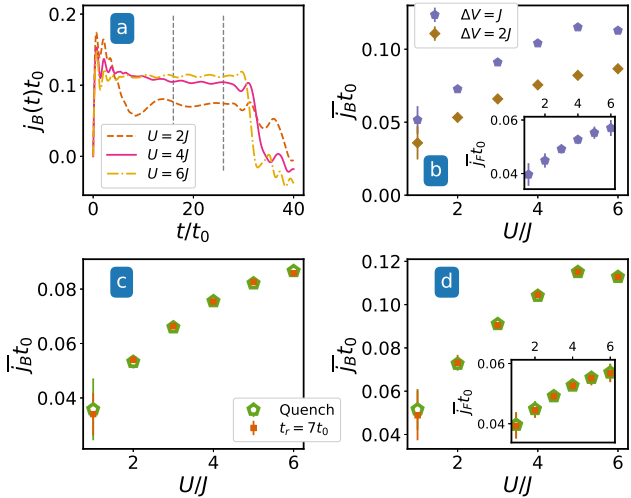}
    \caption{
      Currents in potential-induced transport of the BHM with $L\!=\!60$ sites and $\overline{n}_B\!=\!1/2$.
      (a) Currents at the interface of the potential imbalance versus time for different uniform  b interactions.
      Here $\Delta V\!=\!J$.
      (b) QSSCs from systems with different uniform interactions, showing clear dependence on the initial states.
      The insets show the results for the FHM with $\overline{n}_F\!=\!1/6$ and similar parameters.
      (c) and (d) show the averaged steady-state values from different ramping times when the same potential imbalance is switched on linearly in time.
      $\Delta V\!=\!2J$ in (c) and $\Delta V\!=\!J$ in (d), and all symbols overlap within the error bars for each $U$.
      Memory effects of switching protocols are not resolvable here.
    }
    \label{fig:RampVMemory}
  \end{center}
\end{figure}
\section{Memory Effects in Potential-induced transport}\label{sec:vinduce}
Alternatively, one can use a potential imbalance to drive a current while keeping the interaction uniform.
The time dependent part of the Hamiltonian is an onsite potential
\begin{equation}
  H_{\text{BHM}}(t)=\mathcal{H}_{\text{BHM}}+\sum_iV(i,t)n_i
\end{equation}
for the BHM and
\begin{equation}
  H_{\text{FHM}}(t)=\mathcal{H}_{\text{FHM}}+\sum_{i,\sigma}V(i,t)n_{i,\sigma}
\end{equation}
for the FHM.
$\mathcal{H}_{\text{BHM}}$ and $\mathcal{H}_{\text{FHM}}$ are defined in Eq.\eqref{eq:bhm} and~\eqref{eq:fhm} with $U(i,t)=U$, $\forall i,t$.
Initially, the potential energy is uniform, $V(i,t\leq0)=0,\forall i$, and the system is in the ground state.
A current can be induced by introducing a step-function imbalance in the potential energy so that $V(i\in L,t\geq0)\!=\!\Delta V\!=\!J$ as shown in Fig.~\ref{fig:RampVTransport}.
The QSSC emerges again in potential-induced transport, so we are able to identify the memory effects in this settings.
We mention that the potential bias is applied uniformly on the left half of the system, unlike the constant potential gradient across the entire system as implemented in the study of Bloch oscillations~\cite{Dekorsy:1995ff,Flayac:2011ft}.

A downside of potential-induced transport is that in isolated systems, energy conservation prevents exchange of particles if the potential imbalance is larger than the bandwidth~\cite{Chien:2013ef}.
As a consequence, the system enters a dynamically insulating state if $\Delta V > 4J$.
Therefore, here we only demonstrate two different values of potential imbalance $\Delta V\!=\!J$ and $2J$.

From the values of QSSCs we can determine if there are memory effects of initial states or switching protocols in potential-induced transport.
The results are summarized in Fig.~\ref{fig:RampVMemory}.
The QSSCs clearly depend on initial interaction $U$ and when $U/J$ is small, the QSSC increases with the uniform interaction.
By comparing two different QSSCs with two different interaction but with the same $\Delta V$ in Fig.~\ref{fig:RampVMemory}(b), the memory effects of initial states are clearly identified.
In contrast, Fig.~\ref{fig:RampVMemory}(c)-(d) show that two systems with the same initial uniform interaction $U$ and potential imbalance $\Delta V$ but with different ramping times have the same steady-state values of the QSSC.
In the available parameter space, we found that memory effects of switching protocols are not resolvable in potential-induced transport in lattices, mainly due to the limited potential difference which should be less than the bandwidth.
The same conclusions can be drawn for the FHM and the results are shown in the insets of Fig.~\ref{fig:RampVMemory}(b) and (d).

\section{Experimental implications}\label{sec:exp}
The mass current can be experimentally obtained from the time derivative of the density profile~\cite{Greif:2015wi,Parsons:2016gr,Cheuk:2016tp,Leder:2016wy,Miranda:2014wi}.
In order to observe the quasi-steady states, one may need at least $30$ lattice sites~\cite{Chien:2012ft,Bushong:2005kx}.
The memory effects of initial states can be maximized by using two systems with very different initial interactions so their initial ground states are distinct.
For the memory effects of switching protocols, one has to ramp the interaction imbalance with different time scales, and our results suggest that fermions at half-filling or boson at unity filling will exhibit stronger memory effects.
However, the stronger interaction energy limits the time window for measuring the QSSC and this limitation is also discussed.
The memory effects presented here should survive in the thermodynamic limit when the system size becomes infinity as long as no additional relaxation is introduced.
Since the BHM supports superfluids, dissipation effects such as phase slip~\cite{McKay:2008kt,Abbate2017} may relax the system and eventually wash out the memory effects found here.

Both fermionic and bosonic quantum atomic gases can be trapped in one dimensional (1D) lattices~\cite{Jaksch:2005go,Bloch:2008gl,Morsch:2002dk,Esslinger:2010ex}, and the interactions between atoms can be tune by external magnetic fields or optical means~\cite{Chin:2010kl}.
A spatial interaction imbalance can be produced experimentally by nonuniform magnetic field~\cite{Schunck:2005cf,Chin:2010kl}, or optical control of the atomic collisions~\cite{Clark:2015uj,Fu:2013im,Yamazaki:2010en,Fatemi:2000iy,Enomoto:2008bx,Fukuhara:2009bv,Wu:2012jp}.
Ref.~\cite{Clark:2015uj} has demonstrated a spatial density modulation using an optical Feshbach resonance of bosons, and one may also use inhomogeneous confinement potentials to modulate the interactions in real space~\cite{Stock:2006gw,Moritz:2005bf,Sekh:2012fn,Tiesinga:2000fm}.
The external magnetic field can change the initial interaction~\cite{Chin:2010kl} while the optical control provides an interaction imbalance.
Although it may be possible to use a non-uniform magnetic field to induce interaction imbalance, the challenge is to make the interface separating regions with different interactions narrow compared to the atomic cloud size.
By making the spatially imbalanced interaction time-dependent, one can drive the system out of equilibrium and probe memory effects.
We remark that allowing a finite width of the interaction-imbalance interface does not lead to any qualitative difference in our findings if the width is small, but as temperature increases, the smooth spreading of particle distribution is expected to reduce quantum memory effects.

\section{Conclusion}\label{sec:conclude}
Interaction-induced transport properties are shown to be a versatile platform for studying quantum dynamics, in particular steady-state memory effects.
For dilute atomic gases, the values of QSSC increases (decreases) as a stronger interaction imbalance (initial interaction) is applied to an isolated system.
For bosons with unity filling and fermions with half filling, the behavior is qualitatively the same, and the breakdown of Mott insulating state have been found in both statistics.
In contrast to electronic systems where external relaxation mechanisms such as interactions with impurities and the background ionic lattice bring the systems into a unique equilibrium or steady state, in isolated quantum systems without relaxation long-time memory effects can persist.
Evidence of memory effects in the quasi-steady states induced by interaction imbalance has been presented here for both BHM and FHM.
The memory effects of initial states are quite prevalent, but the memory effects of switching protocols are mostly visible only in lattice interaction-induced transport.
The memory effects discussed here in isolated quantum systems can be understood in the following way: As the states evolve with different but conserved energy, they result in different final states even if the final Hamiltonian is the same.
Similar effect are also discussed in doublon dynamics in both theoretical and experimental studies~\cite{Huber:2009bu,Strohmaier:2010fr,Covey:2016bc,Rausch:2017ck}, where the highly occupied states cannot be easily relaxed due to energy conservation.

Memory effects of steady-state currents in isolated quantum systems can lead to non-constant diffusion coefficients if the kinetic-equation approach is used to describe the steady-state behavior.
Moreover, the steady-state current in quantum systems may find future applications.
For instance, atomic superfluids in ring-shape potential may simulate superconducting devices~\cite{PhysRevLett.111.205301}.
Exploiting memory effects of steady-state currents in isolated systems may lead to alternative designs of quantum devices as proposed in Ref.~\cite{Lai:2016hr}.

\begin{acknowledgments}
We would like to thank Massimiliano Di Ventra and Micheal Zwolak for stimulating discussions.
The tDMRG programs are build upon universal tensor library(Uni10)~\cite{Kao:2015gb}.
Part of the simulations are carried out by Merced Cluster in UC Merced supported by the National Science Foundation (Grant No. ACI-1429783).
\end{acknowledgments}


\begin{thebibliography}{113}
\expandafter\ifx\csname natexlab\endcsname\relax\def\natexlab#1{#1}\fi
\expandafter\ifx\csname bibnamefont\endcsname\relax
  \def\bibnamefont#1{#1}\fi
\expandafter\ifx\csname bibfnamefont\endcsname\relax
  \def\bibfnamefont#1{#1}\fi
\expandafter\ifx\csname citenamefont\endcsname\relax
  \def\citenamefont#1{#1}\fi
\expandafter\ifx\csname url\endcsname\relax
  \def\url#1{\texttt{#1}}\fi
\expandafter\ifx\csname urlprefix\endcsname\relax\def\urlprefix{URL }\fi
\providecommand{\bibinfo}[2]{#2}
\providecommand{\eprint}[2][]{\url{#2}}

\bibitem[{\citenamefont{Schoenlein et~al.}(1987)\citenamefont{Schoenlein, Lin,
  Fujimoto, and Eesley}}]{Schoenlein:1987dg}
\bibinfo{author}{\bibfnamefont{R.~W.} \bibnamefont{Schoenlein}},
  \bibinfo{author}{\bibfnamefont{W.~Z.} \bibnamefont{Lin}},
  \bibinfo{author}{\bibfnamefont{J.~G.} \bibnamefont{Fujimoto}},
  \bibnamefont{and} \bibinfo{author}{\bibfnamefont{G.~L.}
  \bibnamefont{Eesley}}, \bibinfo{journal}{Phys. Rev. Lett.}
  \textbf{\bibinfo{volume}{58}}, \bibinfo{pages}{1680} (\bibinfo{year}{1987}).

\bibitem[{\citenamefont{Roberti et~al.}(1995)\citenamefont{Roberti, Smith, and
  Zhang}}]{Roberti:1995du}
\bibinfo{author}{\bibfnamefont{T.~W.} \bibnamefont{Roberti}},
  \bibinfo{author}{\bibfnamefont{B.~A.} \bibnamefont{Smith}}, \bibnamefont{and}
  \bibinfo{author}{\bibfnamefont{J.~Z.} \bibnamefont{Zhang}},
  \bibinfo{journal}{J. Chem. Phys} \textbf{\bibinfo{volume}{102}},
  \bibinfo{pages}{3860} (\bibinfo{year}{1995}).

\bibitem[{\citenamefont{Bigot et~al.}(1995)\citenamefont{Bigot, Merle, Cregut,
  and Daunois}}]{Bigot:1995he}
\bibinfo{author}{\bibfnamefont{J.~Y.} \bibnamefont{Bigot}},
  \bibinfo{author}{\bibfnamefont{J.~C.} \bibnamefont{Merle}},
  \bibinfo{author}{\bibfnamefont{O.}~\bibnamefont{Cregut}}, \bibnamefont{and}
  \bibinfo{author}{\bibfnamefont{A.}~\bibnamefont{Daunois}},
  \bibinfo{journal}{Phys. Rev. Lett.} \textbf{\bibinfo{volume}{75}},
  \bibinfo{pages}{4702} (\bibinfo{year}{1995}).

\bibitem[{\citenamefont{Bertotti}(1998)}]{bertotti1998hysteresis}
\bibinfo{author}{\bibfnamefont{G.}~\bibnamefont{Bertotti}},
  \emph{\bibinfo{title}{{Hysteresis in Magnetism: For Physicists, Materials
  Scientists, and Engineers}}}, Academic Press series in electromagnetism
  (\bibinfo{publisher}{Academic Press}, \bibinfo{year}{1998}).

\bibitem[{\citenamefont{Jagla}(2017)}]{Jagla:2015vd}
\bibinfo{author}{\bibfnamefont{E.}~\bibnamefont{Jagla}},
  \bibinfo{journal}{Papers in Physics} \textbf{\bibinfo{volume}{9}},
  \bibinfo{pages}{090004} (\bibinfo{year}{2017}).

\bibitem[{\citenamefont{Pershin and Di~Ventra}(2011)}]{Pershin:2011ie}
\bibinfo{author}{\bibfnamefont{Y.~V.} \bibnamefont{Pershin}} \bibnamefont{and}
  \bibinfo{author}{\bibfnamefont{M.}~\bibnamefont{Di~Ventra}},
  \bibinfo{journal}{Adv. Phys.} \textbf{\bibinfo{volume}{60}},
  \bibinfo{pages}{145} (\bibinfo{year}{2011}).

\bibitem[{\citenamefont{Marani et~al.}(2015)\citenamefont{Marani, Gelao, and
  Perri}}]{Marani:2015we}
\bibinfo{author}{\bibfnamefont{R.}~\bibnamefont{Marani}},
  \bibinfo{author}{\bibfnamefont{G.}~\bibnamefont{Gelao}}, \bibnamefont{and}
  \bibinfo{author}{\bibfnamefont{A.~G.} \bibnamefont{Perri}},
  \bibinfo{journal}{IJAET} \textbf{\bibinfo{volume}{8}}, \bibinfo{pages}{294}
  (\bibinfo{year}{2015}).

\bibitem[{\citenamefont{Gilbert et~al.}(2015)\citenamefont{Gilbert, Chern,
  Fore, Lao, Zhang, Nisoli, and Schiffer}}]{Gilbert:2015vx}
\bibinfo{author}{\bibfnamefont{I.}~\bibnamefont{Gilbert}},
  \bibinfo{author}{\bibfnamefont{G.-W.} \bibnamefont{Chern}},
  \bibinfo{author}{\bibfnamefont{B.}~\bibnamefont{Fore}},
  \bibinfo{author}{\bibfnamefont{Y.}~\bibnamefont{Lao}},
  \bibinfo{author}{\bibfnamefont{S.}~\bibnamefont{Zhang}},
  \bibinfo{author}{\bibfnamefont{C.}~\bibnamefont{Nisoli}}, \bibnamefont{and}
  \bibinfo{author}{\bibfnamefont{P.}~\bibnamefont{Schiffer}},
  \bibinfo{journal}{Phys. Rev. B} \textbf{\bibinfo{volume}{92}},
  \bibinfo{pages}{104417} (\bibinfo{year}{2015}).

\bibitem[{\citenamefont{Zhao et~al.}(2017)\citenamefont{Zhao, McLain, Vijande,
  Ferrando, Carr, and Garc{\'\i}a-March}}]{Zhao:2017up}
\bibinfo{author}{\bibfnamefont{X.}~\bibnamefont{Zhao}},
  \bibinfo{author}{\bibfnamefont{M.~A.} \bibnamefont{McLain}},
  \bibinfo{author}{\bibfnamefont{J.}~\bibnamefont{Vijande}},
  \bibinfo{author}{\bibfnamefont{A.}~\bibnamefont{Ferrando}},
  \bibinfo{author}{\bibfnamefont{L.~D.} \bibnamefont{Carr}}, \bibnamefont{and}
  \bibinfo{author}{\bibfnamefont{M.~{\'A}.} \bibnamefont{Garc{\'\i}a-March}},
  \bibinfo{journal}{arXiv:1705.02051} (\bibinfo{year}{2017}).

\bibitem[{\citenamefont{Folcia et~al.}(1987)\citenamefont{Folcia, Tello, and
  P{\'e}rez-Mato}}]{Folcia:1987bs}
\bibinfo{author}{\bibfnamefont{C.~L.} \bibnamefont{Folcia}},
  \bibinfo{author}{\bibfnamefont{M.~J.} \bibnamefont{Tello}}, \bibnamefont{and}
  \bibinfo{author}{\bibfnamefont{J.~M.} \bibnamefont{P{\'e}rez-Mato}},
  \bibinfo{journal}{Phys. Rev. B} \textbf{\bibinfo{volume}{36}},
  \bibinfo{pages}{7181} (\bibinfo{year}{1987}).

\bibitem[{\citenamefont{Aliyev et~al.}(2003)\citenamefont{Aliyev, Babayev,
  Mammadov, Seyidov, and Suleymanov}}]{Aliyev:2003jn}
\bibinfo{author}{\bibfnamefont{V.~P.} \bibnamefont{Aliyev}},
  \bibinfo{author}{\bibfnamefont{S.~S.} \bibnamefont{Babayev}},
  \bibinfo{author}{\bibfnamefont{T.~G.} \bibnamefont{Mammadov}},
  \bibinfo{author}{\bibfnamefont{M.-H.~Y.} \bibnamefont{Seyidov}},
  \bibnamefont{and} \bibinfo{author}{\bibfnamefont{R.~A.}
  \bibnamefont{Suleymanov}}, \bibinfo{journal}{Solid State Communications}
  \textbf{\bibinfo{volume}{128}}, \bibinfo{pages}{25} (\bibinfo{year}{2003}).

\bibitem[{\citenamefont{Marti et~al.}(2014)\citenamefont{Marti, Fina, Frontera,
  Liu, Wadley, He, Paull, Clarkson, Kudrnovsk{\'{y}}, Turek
  et~al.}}]{Marti:2014fl}
\bibinfo{author}{\bibfnamefont{X.}~\bibnamefont{Marti}},
  \bibinfo{author}{\bibfnamefont{I.}~\bibnamefont{Fina}},
  \bibinfo{author}{\bibfnamefont{C.}~\bibnamefont{Frontera}},
  \bibinfo{author}{\bibfnamefont{J.}~\bibnamefont{Liu}},
  \bibinfo{author}{\bibfnamefont{P.}~\bibnamefont{Wadley}},
  \bibinfo{author}{\bibfnamefont{Q.}~\bibnamefont{He}},
  \bibinfo{author}{\bibfnamefont{R.~J.} \bibnamefont{Paull}},
  \bibinfo{author}{\bibfnamefont{J.~D.} \bibnamefont{Clarkson}},
  \bibinfo{author}{\bibfnamefont{J.}~\bibnamefont{Kudrnovsk{\'{y}}}},
  \bibinfo{author}{\bibfnamefont{I.}~\bibnamefont{Turek}},
  \bibnamefont{et~al.}, \bibinfo{journal}{Nat. Mater.}
  \textbf{\bibinfo{volume}{13}}, \bibinfo{pages}{367} (\bibinfo{year}{2014}).

\bibitem[{\citenamefont{Sun et~al.}(2003)\citenamefont{Sun, Salamon, Garnier,
  and Averback}}]{Sun:2003iz}
\bibinfo{author}{\bibfnamefont{Y.}~\bibnamefont{Sun}},
  \bibinfo{author}{\bibfnamefont{M.~B.} \bibnamefont{Salamon}},
  \bibinfo{author}{\bibfnamefont{K.}~\bibnamefont{Garnier}}, \bibnamefont{and}
  \bibinfo{author}{\bibfnamefont{R.~S.} \bibnamefont{Averback}},
  \bibinfo{journal}{Phys. Rev. Lett.} \textbf{\bibinfo{volume}{91}},
  \bibinfo{pages}{167206} (\bibinfo{year}{2003}).

\bibitem[{\citenamefont{Maldacena et~al.}(2016)\citenamefont{Maldacena,
  Shenker, and Stanford}}]{Maldacena:2016gp}
\bibinfo{author}{\bibfnamefont{J.}~\bibnamefont{Maldacena}},
  \bibinfo{author}{\bibfnamefont{S.~H.} \bibnamefont{Shenker}},
  \bibnamefont{and} \bibinfo{author}{\bibfnamefont{D.}~\bibnamefont{Stanford}},
  \bibinfo{journal}{J. High Energ. Phys.} \textbf{\bibinfo{volume}{2016}},
  \bibinfo{pages}{106} (\bibinfo{year}{2016}).

\bibitem[{\citenamefont{Polkovnikov et~al.}(2011)\citenamefont{Polkovnikov,
  Sengupta, Silva, and Vengalattore}}]{Polkovnikov:2011iu}
\bibinfo{author}{\bibfnamefont{A.}~\bibnamefont{Polkovnikov}},
  \bibinfo{author}{\bibfnamefont{K.}~\bibnamefont{Sengupta}},
  \bibinfo{author}{\bibfnamefont{A.}~\bibnamefont{Silva}}, \bibnamefont{and}
  \bibinfo{author}{\bibfnamefont{M.}~\bibnamefont{Vengalattore}},
  \bibinfo{journal}{Rev. Mod. Phys.} \textbf{\bibinfo{volume}{83}},
  \bibinfo{pages}{863} (\bibinfo{year}{2011}).

\bibitem[{\citenamefont{Pasienski et~al.}(2010)\citenamefont{Pasienski, McKay,
  White, and DeMarco}}]{Pasienski:2010bd}
\bibinfo{author}{\bibfnamefont{M.}~\bibnamefont{Pasienski}},
  \bibinfo{author}{\bibfnamefont{D.}~\bibnamefont{McKay}},
  \bibinfo{author}{\bibfnamefont{M.}~\bibnamefont{White}}, \bibnamefont{and}
  \bibinfo{author}{\bibfnamefont{B.}~\bibnamefont{DeMarco}},
  \bibinfo{journal}{Nat. Phys.} \textbf{\bibinfo{volume}{6}},
  \bibinfo{pages}{677} (\bibinfo{year}{2010}).

\bibitem[{\citenamefont{Meldgin et~al.}(2016)\citenamefont{Meldgin, Ray, Russ,
  Chen, Ceperley, and DeMarco}}]{Meldgin:2016cu}
\bibinfo{author}{\bibfnamefont{C.}~\bibnamefont{Meldgin}},
  \bibinfo{author}{\bibfnamefont{U.}~\bibnamefont{Ray}},
  \bibinfo{author}{\bibfnamefont{P.}~\bibnamefont{Russ}},
  \bibinfo{author}{\bibfnamefont{D.}~\bibnamefont{Chen}},
  \bibinfo{author}{\bibfnamefont{D.~M.} \bibnamefont{Ceperley}},
  \bibnamefont{and} \bibinfo{author}{\bibfnamefont{B.}~\bibnamefont{DeMarco}},
  \bibinfo{journal}{Nat. Phys.}  (\bibinfo{year}{2016}).

\bibitem[{\citenamefont{Kondov et~al.}(2015)\citenamefont{Kondov, McGehee, Xu,
  and DeMarco}}]{Kondov:2015ce}
\bibinfo{author}{\bibfnamefont{S.~S.} \bibnamefont{Kondov}},
  \bibinfo{author}{\bibfnamefont{W.~R.} \bibnamefont{McGehee}},
  \bibinfo{author}{\bibfnamefont{W.}~\bibnamefont{Xu}}, \bibnamefont{and}
  \bibinfo{author}{\bibfnamefont{B.}~\bibnamefont{DeMarco}},
  \bibinfo{journal}{Phys. Rev. Lett.} \textbf{\bibinfo{volume}{114}},
  \bibinfo{pages}{083002} (\bibinfo{year}{2015}).

\bibitem[{\citenamefont{Fischer et~al.}(2016)\citenamefont{Fischer, Maksymenko,
  and Altman}}]{Fischer:2015ve}
\bibinfo{author}{\bibfnamefont{M.~H.} \bibnamefont{Fischer}},
  \bibinfo{author}{\bibfnamefont{M.}~\bibnamefont{Maksymenko}},
  \bibnamefont{and} \bibinfo{author}{\bibfnamefont{E.}~\bibnamefont{Altman}},
  \bibinfo{journal}{Phys. Rev. Lett.} \textbf{\bibinfo{volume}{116}},
  \bibinfo{pages}{160401} (\bibinfo{year}{2016}).

\bibitem[{\citenamefont{Schreiber et~al.}(2015)\citenamefont{Schreiber,
  Hodgman, Bordia, L{\"u}schen, Fischer, Vosk, Altman, Schneider, and
  Bloch}}]{Schreiber:2015jt}
\bibinfo{author}{\bibfnamefont{M.}~\bibnamefont{Schreiber}},
  \bibinfo{author}{\bibfnamefont{S.~S.} \bibnamefont{Hodgman}},
  \bibinfo{author}{\bibfnamefont{P.}~\bibnamefont{Bordia}},
  \bibinfo{author}{\bibfnamefont{H.~P.} \bibnamefont{L{\"u}schen}},
  \bibinfo{author}{\bibfnamefont{M.~H.} \bibnamefont{Fischer}},
  \bibinfo{author}{\bibfnamefont{R.}~\bibnamefont{Vosk}},
  \bibinfo{author}{\bibfnamefont{E.}~\bibnamefont{Altman}},
  \bibinfo{author}{\bibfnamefont{U.}~\bibnamefont{Schneider}},
  \bibnamefont{and} \bibinfo{author}{\bibfnamefont{I.}~\bibnamefont{Bloch}},
  \bibinfo{journal}{Science} \textbf{\bibinfo{volume}{349}},
  \bibinfo{pages}{842} (\bibinfo{year}{2015}).

\bibitem[{\citenamefont{Smith et~al.}(2016)\citenamefont{Smith, Lee, Richerme,
  Neyenhuis, Hess, Hauke, Heyl, Huse, and Monroe}}]{Smith:2015uo}
\bibinfo{author}{\bibfnamefont{J.}~\bibnamefont{Smith}},
  \bibinfo{author}{\bibfnamefont{A.}~\bibnamefont{Lee}},
  \bibinfo{author}{\bibfnamefont{P.}~\bibnamefont{Richerme}},
  \bibinfo{author}{\bibfnamefont{B.}~\bibnamefont{Neyenhuis}},
  \bibinfo{author}{\bibfnamefont{P.~W.} \bibnamefont{Hess}},
  \bibinfo{author}{\bibfnamefont{P.}~\bibnamefont{Hauke}},
  \bibinfo{author}{\bibfnamefont{M.}~\bibnamefont{Heyl}},
  \bibinfo{author}{\bibfnamefont{D.~A.} \bibnamefont{Huse}}, \bibnamefont{and}
  \bibinfo{author}{\bibfnamefont{C.}~\bibnamefont{Monroe}},
  \bibinfo{journal}{Nat. Phys.} \textbf{\bibinfo{volume}{12}},
  \bibinfo{pages}{907} (\bibinfo{year}{2016}).

\bibitem[{\citenamefont{Tsuji et~al.}(2017)\citenamefont{Tsuji, Werner, and
  Ueda}}]{Tsuji:2017hm}
\bibinfo{author}{\bibfnamefont{N.}~\bibnamefont{Tsuji}},
  \bibinfo{author}{\bibfnamefont{P.}~\bibnamefont{Werner}}, \bibnamefont{and}
  \bibinfo{author}{\bibfnamefont{M.}~\bibnamefont{Ueda}},
  \bibinfo{journal}{Phys. Rev. A} \textbf{\bibinfo{volume}{95}},
  \bibinfo{pages}{011601} (\bibinfo{year}{2017}).

\bibitem[{\citenamefont{D{\'o}ra and Moessner}(2017)}]{Dora:2017go}
\bibinfo{author}{\bibfnamefont{B.}~\bibnamefont{D{\'o}ra}} \bibnamefont{and}
  \bibinfo{author}{\bibfnamefont{R.}~\bibnamefont{Moessner}},
  \bibinfo{journal}{Phys. Rev. Lett.} \textbf{\bibinfo{volume}{119}},
  \bibinfo{pages}{026802} (\bibinfo{year}{2017}).

\bibitem[{\citenamefont{Haehl et~al.}(2017)\citenamefont{Haehl, Loganayagam,
  Narayan, and Rangamani}}]{Haehl:2017ui}
\bibinfo{author}{\bibfnamefont{F.~M.} \bibnamefont{Haehl}},
  \bibinfo{author}{\bibfnamefont{R.}~\bibnamefont{Loganayagam}},
  \bibinfo{author}{\bibfnamefont{P.}~\bibnamefont{Narayan}}, \bibnamefont{and}
  \bibinfo{author}{\bibfnamefont{M.}~\bibnamefont{Rangamani}},
  \bibinfo{journal}{arXiv:1701.02820} (\bibinfo{year}{2017}).

\bibitem[{\citenamefont{Yao et~al.}(2016)\citenamefont{Yao, Grusdt, Swingle,
  Lukin, Stamper-Kurn, Moore, and Demler}}]{Yao:2016tb}
\bibinfo{author}{\bibfnamefont{N.~Y.} \bibnamefont{Yao}},
  \bibinfo{author}{\bibfnamefont{F.}~\bibnamefont{Grusdt}},
  \bibinfo{author}{\bibfnamefont{B.}~\bibnamefont{Swingle}},
  \bibinfo{author}{\bibfnamefont{M.~D.} \bibnamefont{Lukin}},
  \bibinfo{author}{\bibfnamefont{D.~M.} \bibnamefont{Stamper-Kurn}},
  \bibinfo{author}{\bibfnamefont{J.~E.} \bibnamefont{Moore}}, \bibnamefont{and}
  \bibinfo{author}{\bibfnamefont{E.~A.} \bibnamefont{Demler}},
  \bibinfo{journal}{arXiv:1607.01801} (\bibinfo{year}{2016}).

\bibitem[{\citenamefont{Chien et~al.}(2015)\citenamefont{Chien, Peotta, and
  Di~Ventra}}]{Chien:2015kc}
\bibinfo{author}{\bibfnamefont{C.-C.} \bibnamefont{Chien}},
  \bibinfo{author}{\bibfnamefont{S.}~\bibnamefont{Peotta}}, \bibnamefont{and}
  \bibinfo{author}{\bibfnamefont{M.}~\bibnamefont{Di~Ventra}},
  \bibinfo{journal}{Nat. Phys.} \textbf{\bibinfo{volume}{11}},
  \bibinfo{pages}{998} (\bibinfo{year}{2015}).

\bibitem[{\citenamefont{Eckel et~al.}(2014)\citenamefont{Eckel, Lee,
  Jendrzejewski, Murray, Clark, Lobb, Phillips, Edwards, and
  Campbell}}]{Eckel:2014gf}
\bibinfo{author}{\bibfnamefont{S.}~\bibnamefont{Eckel}},
  \bibinfo{author}{\bibfnamefont{J.~G.} \bibnamefont{Lee}},
  \bibinfo{author}{\bibfnamefont{F.}~\bibnamefont{Jendrzejewski}},
  \bibinfo{author}{\bibfnamefont{N.}~\bibnamefont{Murray}},
  \bibinfo{author}{\bibfnamefont{C.~W.} \bibnamefont{Clark}},
  \bibinfo{author}{\bibfnamefont{C.~J.} \bibnamefont{Lobb}},
  \bibinfo{author}{\bibfnamefont{W.~D.} \bibnamefont{Phillips}},
  \bibinfo{author}{\bibfnamefont{M.}~\bibnamefont{Edwards}}, \bibnamefont{and}
  \bibinfo{author}{\bibfnamefont{G.~K.} \bibnamefont{Campbell}},
  \bibinfo{journal}{Nature (London)} \textbf{\bibinfo{volume}{506}},
  \bibinfo{pages}{200} (\bibinfo{year}{2014}).

\bibitem[{\citenamefont{Metcalf et~al.}(2016)\citenamefont{Metcalf, Lai, and
  Chien}}]{Metcalf:2016dm}
\bibinfo{author}{\bibfnamefont{M.}~\bibnamefont{Metcalf}},
  \bibinfo{author}{\bibfnamefont{C.-Y.} \bibnamefont{Lai}}, \bibnamefont{and}
  \bibinfo{author}{\bibfnamefont{C.-C.} \bibnamefont{Chien}},
  \bibinfo{journal}{Phys. Rev. A} \textbf{\bibinfo{volume}{93}},
  \bibinfo{pages}{053617} (\bibinfo{year}{2016}).

\bibitem[{\citenamefont{Costa-Filho et~al.}(2017)\citenamefont{Costa-Filho,
  Lima, Paiva, Soares, Morgado, Franco, and Soares-Pinto}}]{PhysRevA.95.052126}
\bibinfo{author}{\bibfnamefont{J.~I.} \bibnamefont{Costa-Filho}},
  \bibinfo{author}{\bibfnamefont{R.~B.~B.} \bibnamefont{Lima}},
  \bibinfo{author}{\bibfnamefont{R.~R.} \bibnamefont{Paiva}},
  \bibinfo{author}{\bibfnamefont{P.~M.} \bibnamefont{Soares}},
  \bibinfo{author}{\bibfnamefont{W.~A.~M.} \bibnamefont{Morgado}},
  \bibinfo{author}{\bibfnamefont{R.~L.} \bibnamefont{Franco}},
  \bibnamefont{and} \bibinfo{author}{\bibfnamefont{D.~O.}
  \bibnamefont{Soares-Pinto}}, \bibinfo{journal}{Phys. Rev. A}
  \textbf{\bibinfo{volume}{95}}, \bibinfo{pages}{052126}
  (\bibinfo{year}{2017}).

\bibitem[{\citenamefont{Man et~al.}(2015)\citenamefont{Man, Xia, and
  Lo~Franco}}]{PhysRevA.92.012315}
\bibinfo{author}{\bibfnamefont{Z.-X.} \bibnamefont{Man}},
  \bibinfo{author}{\bibfnamefont{Y.-J.} \bibnamefont{Xia}}, \bibnamefont{and}
  \bibinfo{author}{\bibfnamefont{R.}~\bibnamefont{Lo~Franco}},
  \bibinfo{journal}{Phys. Rev. A} \textbf{\bibinfo{volume}{92}},
  \bibinfo{pages}{012315} (\bibinfo{year}{2015}).

\bibitem[{\citenamefont{Cornean et~al.}(2014)\citenamefont{Cornean, Jensen, and
  Nenciu}}]{Cornean:2013kz}
\bibinfo{author}{\bibfnamefont{H.~D.} \bibnamefont{Cornean}},
  \bibinfo{author}{\bibfnamefont{A.}~\bibnamefont{Jensen}}, \bibnamefont{and}
  \bibinfo{author}{\bibfnamefont{G.}~\bibnamefont{Nenciu}},
  \bibinfo{journal}{Ann. Henri Poincar{\'e}} \textbf{\bibinfo{volume}{15}},
  \bibinfo{pages}{1919} (\bibinfo{year}{2014}).

\bibitem[{\citenamefont{Chien et~al.}(2013)\citenamefont{Chien, Gruss,
  Di~Ventra, and Zwolak}}]{Chien:2013ef}
\bibinfo{author}{\bibfnamefont{C.-C.} \bibnamefont{Chien}},
  \bibinfo{author}{\bibfnamefont{D.}~\bibnamefont{Gruss}},
  \bibinfo{author}{\bibfnamefont{M.}~\bibnamefont{Di~Ventra}},
  \bibnamefont{and} \bibinfo{author}{\bibfnamefont{M.}~\bibnamefont{Zwolak}},
  \bibinfo{journal}{New J. Phys.} \textbf{\bibinfo{volume}{15}},
  \bibinfo{pages}{063026} (\bibinfo{year}{2013}).

\bibitem[{\citenamefont{Chien and Di~Ventra}(2013)}]{Chien:2013en}
\bibinfo{author}{\bibfnamefont{C.-C.} \bibnamefont{Chien}} \bibnamefont{and}
  \bibinfo{author}{\bibfnamefont{M.}~\bibnamefont{Di~Ventra}},
  \bibinfo{journal}{Phys. Rev. A} \textbf{\bibinfo{volume}{87}},
  \bibinfo{pages}{023609} (\bibinfo{year}{2013}).

\bibitem[{\citenamefont{Chien et~al.}(2014)\citenamefont{Chien, Di~Ventra, and
  Zwolak}}]{Chien:2014uf}
\bibinfo{author}{\bibfnamefont{C.-C.} \bibnamefont{Chien}},
  \bibinfo{author}{\bibfnamefont{M.}~\bibnamefont{Di~Ventra}},
  \bibnamefont{and} \bibinfo{author}{\bibfnamefont{M.}~\bibnamefont{Zwolak}},
  \bibinfo{journal}{Phys. Rev. A} \textbf{\bibinfo{volume}{90}},
  \bibinfo{pages}{023624} (\bibinfo{year}{2014}).

\bibitem[{\citenamefont{Lai and Chien}(2016{\natexlab{a}})}]{Lai:2016hr}
\bibinfo{author}{\bibfnamefont{C.-Y.} \bibnamefont{Lai}} \bibnamefont{and}
  \bibinfo{author}{\bibfnamefont{C.-C.} \bibnamefont{Chien}},
  \bibinfo{journal}{Phys. Rev. Applied} \textbf{\bibinfo{volume}{5}},
  \bibinfo{pages}{034001} (\bibinfo{year}{2016}{\natexlab{a}}).

\bibitem[{\citenamefont{He and Chien}(2016)}]{He:2016fc}
\bibinfo{author}{\bibfnamefont{Y.}~\bibnamefont{He}} \bibnamefont{and}
  \bibinfo{author}{\bibfnamefont{C.-C.} \bibnamefont{Chien}},
  \bibinfo{journal}{Phys. Rev. B} \textbf{\bibinfo{volume}{94}},
  \bibinfo{pages}{024308} (\bibinfo{year}{2016}).

\bibitem[{\citenamefont{Metcalf et~al.}(2017)\citenamefont{Metcalf, Lai,
  Wright, and Chien}}]{Metcalf:2017vya}
\bibinfo{author}{\bibfnamefont{M.}~\bibnamefont{Metcalf}},
  \bibinfo{author}{\bibfnamefont{C.-Y.} \bibnamefont{Lai}},
  \bibinfo{author}{\bibfnamefont{K.}~\bibnamefont{Wright}}, \bibnamefont{and}
  \bibinfo{author}{\bibfnamefont{C.-C.} \bibnamefont{Chien}},
  \bibinfo{journal}{Europhysics Lett. (EPL)} \textbf{\bibinfo{volume}{118}},
  \bibinfo{pages}{56004} (\bibinfo{year}{2017}).

\bibitem[{\citenamefont{Wang et~al.}(2015)\citenamefont{Wang, Xu, Wang, and
  Wu}}]{DaWang:2015hr}
\bibinfo{author}{\bibfnamefont{D.}~\bibnamefont{Wang}},
  \bibinfo{author}{\bibfnamefont{S.}~\bibnamefont{Xu}},
  \bibinfo{author}{\bibfnamefont{Y.}~\bibnamefont{Wang}}, \bibnamefont{and}
  \bibinfo{author}{\bibfnamefont{C.}~\bibnamefont{Wu}}, \bibinfo{journal}{Phys.
  Rev. B} \textbf{\bibinfo{volume}{91}}, \bibinfo{pages}{115118}
  (\bibinfo{year}{2015}).

\bibitem[{\citenamefont{Chung et~al.}(2013)\citenamefont{Chung, Jhu, Chen, and
  Mou}}]{Chung:2013ce}
\bibinfo{author}{\bibfnamefont{M.-C.} \bibnamefont{Chung}},
  \bibinfo{author}{\bibfnamefont{Y.-H.} \bibnamefont{Jhu}},
  \bibinfo{author}{\bibfnamefont{P.}~\bibnamefont{Chen}}, \bibnamefont{and}
  \bibinfo{author}{\bibfnamefont{C.-Y.} \bibnamefont{Mou}},
  \bibinfo{journal}{J. Phys.: Condens. Matter} \textbf{\bibinfo{volume}{25}},
  \bibinfo{pages}{285601} (\bibinfo{year}{2013}).

\bibitem[{\citenamefont{Rigol et~al.}(2006)\citenamefont{Rigol, Muramatsu, and
  Olshanii}}]{Rigol:2006df}
\bibinfo{author}{\bibfnamefont{M.}~\bibnamefont{Rigol}},
  \bibinfo{author}{\bibfnamefont{A.}~\bibnamefont{Muramatsu}},
  \bibnamefont{and} \bibinfo{author}{\bibfnamefont{M.}~\bibnamefont{Olshanii}},
  \bibinfo{journal}{Phys. Rev. A} \textbf{\bibinfo{volume}{74}},
  \bibinfo{pages}{053616} (\bibinfo{year}{2006}).

\bibitem[{\citenamefont{Rigol et~al.}(2007)\citenamefont{Rigol, Dunjko,
  Yurovsky, and Olshanii}}]{Rigol:2007bm}
\bibinfo{author}{\bibfnamefont{M.}~\bibnamefont{Rigol}},
  \bibinfo{author}{\bibfnamefont{V.}~\bibnamefont{Dunjko}},
  \bibinfo{author}{\bibfnamefont{V.}~\bibnamefont{Yurovsky}}, \bibnamefont{and}
  \bibinfo{author}{\bibfnamefont{M.}~\bibnamefont{Olshanii}},
  \bibinfo{journal}{Phys. Rev. Lett.} \textbf{\bibinfo{volume}{98}},
  \bibinfo{pages}{050405} (\bibinfo{year}{2007}).

\bibitem[{\citenamefont{Chung et~al.}(2012)\citenamefont{Chung, Iucci, and
  Cazalilla}}]{Chung:2012bl}
\bibinfo{author}{\bibfnamefont{M.-C.} \bibnamefont{Chung}},
  \bibinfo{author}{\bibfnamefont{A.}~\bibnamefont{Iucci}}, \bibnamefont{and}
  \bibinfo{author}{\bibfnamefont{M.~A.} \bibnamefont{Cazalilla}},
  \bibinfo{journal}{New J. Phys.} \textbf{\bibinfo{volume}{14}},
  \bibinfo{pages}{075013} (\bibinfo{year}{2012}).

\bibitem[{\citenamefont{Perfetto et~al.}(2010)\citenamefont{Perfetto,
  Stefanucci, and Cini}}]{Perfetto:2010ih}
\bibinfo{author}{\bibfnamefont{E.}~\bibnamefont{Perfetto}},
  \bibinfo{author}{\bibfnamefont{G.}~\bibnamefont{Stefanucci}},
  \bibnamefont{and} \bibinfo{author}{\bibfnamefont{M.}~\bibnamefont{Cini}},
  \bibinfo{journal}{Phys. Rev. Lett.} \textbf{\bibinfo{volume}{105}},
  \bibinfo{pages}{156802} (\bibinfo{year}{2010}).

\bibitem[{\citenamefont{White}(1992)}]{White:1992ie}
\bibinfo{author}{\bibfnamefont{S.~R.} \bibnamefont{White}},
  \bibinfo{journal}{Phys. Rev. Lett.} \textbf{\bibinfo{volume}{69}},
  \bibinfo{pages}{2863} (\bibinfo{year}{1992}).

\bibitem[{\citenamefont{White}(1993)}]{White:1993fb}
\bibinfo{author}{\bibfnamefont{S.~R.} \bibnamefont{White}},
  \bibinfo{journal}{Phys. Rev. B} \textbf{\bibinfo{volume}{48}},
  \bibinfo{pages}{10345} (\bibinfo{year}{1993}).

\bibitem[{\citenamefont{{\"O}stlund and Rommer}(1995)}]{Ostlund:1995fx}
\bibinfo{author}{\bibfnamefont{S.}~\bibnamefont{{\"O}stlund}} \bibnamefont{and}
  \bibinfo{author}{\bibfnamefont{S.}~\bibnamefont{Rommer}},
  \bibinfo{journal}{Phys. Rev. Lett.} \textbf{\bibinfo{volume}{75}},
  \bibinfo{pages}{3537} (\bibinfo{year}{1995}).

\bibitem[{\citenamefont{Schollw{\"o}ck}(2005)}]{Schollwock:2005jv}
\bibinfo{author}{\bibfnamefont{U.}~\bibnamefont{Schollw{\"o}ck}},
  \bibinfo{journal}{Rev. Mod. Phys.} \textbf{\bibinfo{volume}{77}},
  \bibinfo{pages}{259} (\bibinfo{year}{2005}).

\bibitem[{\citenamefont{McCulloch}(2007)}]{McCulloch:2007gi}
\bibinfo{author}{\bibfnamefont{I.~P.} \bibnamefont{McCulloch}},
  \bibinfo{journal}{J. Stat. Mech.} \textbf{\bibinfo{volume}{2007}},
  \bibinfo{pages}{P10014} (\bibinfo{year}{2007}).

\bibitem[{\citenamefont{Schollw{\"o}ck}(2011)}]{Schollwock:2011gl}
\bibinfo{author}{\bibfnamefont{U.}~\bibnamefont{Schollw{\"o}ck}},
  \bibinfo{journal}{Ann. Phys.} \textbf{\bibinfo{volume}{326}},
  \bibinfo{pages}{96} (\bibinfo{year}{2011}).

\bibitem[{\citenamefont{White and Feiguin}(2004)}]{White:2004fd}
\bibinfo{author}{\bibfnamefont{S.~R.} \bibnamefont{White}} \bibnamefont{and}
  \bibinfo{author}{\bibfnamefont{A.~E.} \bibnamefont{Feiguin}},
  \bibinfo{journal}{Phys. Rev. Lett.} \textbf{\bibinfo{volume}{93}},
  \bibinfo{pages}{076401} (\bibinfo{year}{2004}).

\bibitem[{\citenamefont{Lai et~al.}(2008)\citenamefont{Lai, Hung, Mou, and
  Chen}}]{Lai:2008ez}
\bibinfo{author}{\bibfnamefont{C.-Y.} \bibnamefont{Lai}},
  \bibinfo{author}{\bibfnamefont{J.-T.} \bibnamefont{Hung}},
  \bibinfo{author}{\bibfnamefont{C.-Y.} \bibnamefont{Mou}}, \bibnamefont{and}
  \bibinfo{author}{\bibfnamefont{P.}~\bibnamefont{Chen}},
  \bibinfo{journal}{Phys. Rev. B} \textbf{\bibinfo{volume}{77}},
  \bibinfo{pages}{205419} (\bibinfo{year}{2008}).

\bibitem[{\citenamefont{Vidal}(2004)}]{Vidal:2004jc}
\bibinfo{author}{\bibfnamefont{G.}~\bibnamefont{Vidal}},
  \bibinfo{journal}{Phys. Rev. Lett.} \textbf{\bibinfo{volume}{93}},
  \bibinfo{pages}{040502} (\bibinfo{year}{2004}).

\bibitem[{\citenamefont{Kamenev}(2011)}]{kamenev2011field}
\bibinfo{author}{\bibfnamefont{A.}~\bibnamefont{Kamenev}},
  \emph{\bibinfo{title}{{Field Theory of Non-Equilibrium Systems}}}
  (\bibinfo{publisher}{Cambridge University Press}, \bibinfo{year}{2011}).

\bibitem[{\citenamefont{Di~Ventra}(2010)}]{DiVentra:2010ks}
\bibinfo{author}{\bibfnamefont{M.}~\bibnamefont{Di~Ventra}},
  \emph{\bibinfo{title}{{Electrical Transport in Nanoscale Systems}}}
  (\bibinfo{publisher}{Cambridge University Press},
  \bibinfo{address}{Cambridge}, \bibinfo{year}{2010}).

\bibitem[{\citenamefont{Chien and Di~Ventra}(2012)}]{Chien:2012cv}
\bibinfo{author}{\bibfnamefont{C.-C.} \bibnamefont{Chien}} \bibnamefont{and}
  \bibinfo{author}{\bibfnamefont{M.}~\bibnamefont{Di~Ventra}},
  \bibinfo{journal}{Europhys. Lett.} \textbf{\bibinfo{volume}{99}},
  \bibinfo{pages}{40003} (\bibinfo{year}{2012}).

\bibitem[{\citenamefont{Di~Ventra and Todorov}(2004)}]{DiVentra:2004bx}
\bibinfo{author}{\bibfnamefont{M.}~\bibnamefont{Di~Ventra}} \bibnamefont{and}
  \bibinfo{author}{\bibfnamefont{T.~N.} \bibnamefont{Todorov}},
  \bibinfo{journal}{J. Phys.: Condens. Matter} \textbf{\bibinfo{volume}{16}},
  \bibinfo{pages}{8025} (\bibinfo{year}{2004}).

\bibitem[{\citenamefont{Labouvie et~al.}(2015)\citenamefont{Labouvie, Santra,
  Heun, Wimberger, and Ott}}]{Labouvie:2015dx}
\bibinfo{author}{\bibfnamefont{R.}~\bibnamefont{Labouvie}},
  \bibinfo{author}{\bibfnamefont{B.}~\bibnamefont{Santra}},
  \bibinfo{author}{\bibfnamefont{S.}~\bibnamefont{Heun}},
  \bibinfo{author}{\bibfnamefont{S.}~\bibnamefont{Wimberger}},
  \bibnamefont{and} \bibinfo{author}{\bibfnamefont{H.}~\bibnamefont{Ott}},
  \bibinfo{journal}{Phys. Rev. Lett.} \textbf{\bibinfo{volume}{115}},
  \bibinfo{pages}{050601} (\bibinfo{year}{2015}).

\bibitem[{\citenamefont{Lai and Chien}(2016{\natexlab{b}})}]{Lai:2016kh}
\bibinfo{author}{\bibfnamefont{C.-Y.} \bibnamefont{Lai}} \bibnamefont{and}
  \bibinfo{author}{\bibfnamefont{C.-C.} \bibnamefont{Chien}},
  \bibinfo{journal}{Sci. Rep.} \textbf{\bibinfo{volume}{6}},
  \bibinfo{pages}{37256} (\bibinfo{year}{2016}{\natexlab{b}}).

\bibitem[{\citenamefont{Oka et~al.}(2003)\citenamefont{Oka, Arita, and
  Aoki}}]{Oka:2003hr}
\bibinfo{author}{\bibfnamefont{T.}~\bibnamefont{Oka}},
  \bibinfo{author}{\bibfnamefont{R.}~\bibnamefont{Arita}}, \bibnamefont{and}
  \bibinfo{author}{\bibfnamefont{H.}~\bibnamefont{Aoki}},
  \bibinfo{journal}{Phys. Rev. Lett.} \textbf{\bibinfo{volume}{91}},
  \bibinfo{pages}{066406} (\bibinfo{year}{2003}).

\bibitem[{\citenamefont{Oka and Aoki}(2005)}]{Oka:2005dk}
\bibinfo{author}{\bibfnamefont{T.}~\bibnamefont{Oka}} \bibnamefont{and}
  \bibinfo{author}{\bibfnamefont{H.}~\bibnamefont{Aoki}},
  \bibinfo{journal}{Phys. Rev. Lett.} \textbf{\bibinfo{volume}{95}},
  \bibinfo{pages}{137601} (\bibinfo{year}{2005}).

\bibitem[{\citenamefont{Heidrich-Meisner
  et~al.}(2010)\citenamefont{Heidrich-Meisner, Gonz{\'a}lez, Al-Hassanieh,
  Feiguin, Rozenberg, and Dagotto}}]{HeidrichMeisner:2010fr}
\bibinfo{author}{\bibfnamefont{F.}~\bibnamefont{Heidrich-Meisner}},
  \bibinfo{author}{\bibfnamefont{I.}~\bibnamefont{Gonz{\'a}lez}},
  \bibinfo{author}{\bibfnamefont{K.~A.} \bibnamefont{Al-Hassanieh}},
  \bibinfo{author}{\bibfnamefont{A.~E.} \bibnamefont{Feiguin}},
  \bibinfo{author}{\bibfnamefont{M.~J.} \bibnamefont{Rozenberg}},
  \bibnamefont{and} \bibinfo{author}{\bibfnamefont{E.}~\bibnamefont{Dagotto}},
  \bibinfo{journal}{Phys. Rev. B} \textbf{\bibinfo{volume}{82}},
  \bibinfo{pages}{205110} (\bibinfo{year}{2010}).

\bibitem[{\citenamefont{K{\"u}hner et~al.}(2000)\citenamefont{K{\"u}hner,
  White, and Monien}}]{Kuhner:2000tg}
\bibinfo{author}{\bibfnamefont{T.D.}~\bibnamefont{K{\"u}hner}},
  \bibinfo{author}{\bibfnamefont{S.~R.} \bibnamefont{White}}, \bibnamefont{and}
  \bibinfo{author}{\bibfnamefont{H.}~\bibnamefont{Monien}},
  \bibinfo{journal}{Phys. Rev. B} \textbf{\bibinfo{volume}{61}},
  \bibinfo{pages}{12474} (\bibinfo{year}{2000}).

\bibitem[{\citenamefont{Zakrzewski and Delande}(2008)}]{Zakrzewski:2008gs}
\bibinfo{author}{\bibfnamefont{J.}~\bibnamefont{Zakrzewski}} \bibnamefont{and}
  \bibinfo{author}{\bibfnamefont{D.}~\bibnamefont{Delande}},
  \bibinfo{journal}{AIP Conference Proceedings}
  \textbf{\bibinfo{volume}{1076}}, \bibinfo{pages}{292} (\bibinfo{year}{2008}).

\bibitem[{\citenamefont{Conwell}(2008)}]{Conwell:2008kf}
\bibinfo{author}{\bibfnamefont{E.~M.} \bibnamefont{Conwell}},
  \bibinfo{journal}{Physics Today} \textbf{\bibinfo{volume}{23}},
  \bibinfo{pages}{35} (\bibinfo{year}{2008}).

\bibitem[{\citenamefont{Oka and Aoki}(2010)}]{Oka:2010ku}
\bibinfo{author}{\bibfnamefont{T.}~\bibnamefont{Oka}} \bibnamefont{and}
  \bibinfo{author}{\bibfnamefont{H.}~\bibnamefont{Aoki}},
  \bibinfo{journal}{Phys. Rev. B} \textbf{\bibinfo{volume}{81}},
  \bibinfo{pages}{033103} (\bibinfo{year}{2010}).

\bibitem[{\citenamefont{Lieb and Wu}(2003)}]{Lieb:2003hl}
\bibinfo{author}{\bibfnamefont{E.~H.} \bibnamefont{Lieb}} \bibnamefont{and}
  \bibinfo{author}{\bibfnamefont{F.~Y.} \bibnamefont{Wu}},
  \bibinfo{journal}{Physica A: Statistical Mechanics and its Applications}
  \textbf{\bibinfo{volume}{321}}, \bibinfo{pages}{1} (\bibinfo{year}{2003}).

\bibitem[{\citenamefont{Grandy~Jr}(2008)}]{2008entropy}
\bibinfo{author}{\bibfnamefont{W.~T.} \bibnamefont{Grandy~Jr}},
  \emph{\bibinfo{title}{{Entropy and the Time Evolution of Macroscopic
  Systems}}}, International Series of Monographs on Physics
  (\bibinfo{publisher}{OUP Oxford}, \bibinfo{year}{2008}).

\bibitem[{\citenamefont{Altland and Simons}(2010)}]{altland2010condensed}
\bibinfo{author}{\bibfnamefont{A.}~\bibnamefont{Altland}} \bibnamefont{and}
  \bibinfo{author}{\bibfnamefont{B.~D.} \bibnamefont{Simons}},
  \emph{\bibinfo{title}{{Condensed Matter Field Theory}}}, Cambridge books
  online (\bibinfo{publisher}{Cambridge University Press},
  \bibinfo{year}{2010}).

\bibitem[{\citenamefont{Gross}(1961)}]{Gross:1961bx}
\bibinfo{author}{\bibfnamefont{E.~P.} \bibnamefont{Gross}},
  \bibinfo{journal}{Nuovo Cimento} \textbf{\bibinfo{volume}{20}},
  \bibinfo{pages}{454} (\bibinfo{year}{1961}).

\bibitem[{\citenamefont{Pitaevsk}(1961)}]{Pitaevsk:fe}
\bibinfo{author}{\bibfnamefont{L.~P.} \bibnamefont{Pitaevsk}},
  \bibinfo{journal}{Soviet Physics JETP-USSR} \textbf{\bibinfo{volume}{13}},
  \bibinfo{pages}{451} (\bibinfo{year}{1961}).

\bibitem[{\citenamefont{Vudragovi{\'c}
  et~al.}(2012)\citenamefont{Vudragovi{\'c}, Vidanovi{\'c}, Bala{\v z},
  Muruganandam, and Adhikari}}]{Vudragovic:2012jz}
\bibinfo{author}{\bibfnamefont{D.}~\bibnamefont{Vudragovi{\'c}}},
  \bibinfo{author}{\bibfnamefont{I.}~\bibnamefont{Vidanovi{\'c}}},
  \bibinfo{author}{\bibfnamefont{A.}~\bibnamefont{Bala{\v z}}},
  \bibinfo{author}{\bibfnamefont{P.}~\bibnamefont{Muruganandam}},
  \bibnamefont{and} \bibinfo{author}{\bibfnamefont{S.~K.}
  \bibnamefont{Adhikari}}, \bibinfo{journal}{Comput. Phys. Commun.}
  \textbf{\bibinfo{volume}{183}}, \bibinfo{pages}{2021} (\bibinfo{year}{2012}).

\bibitem[{\citenamefont{Pethick and Smith}(2010)}]{Pethick:2010gy}
\bibinfo{author}{\bibfnamefont{C.~J.} \bibnamefont{Pethick}} \bibnamefont{and}
  \bibinfo{author}{\bibfnamefont{H.}~\bibnamefont{Smith}},
  \emph{\bibinfo{title}{{Bose{\textendash}Einstein Condensation in Dilute
  Gases}}} (\bibinfo{publisher}{Cambridge University Press},
  \bibinfo{address}{Cambridge}, \bibinfo{year}{2010}), \bibinfo{edition}{2nd}
  ed.

\bibitem[{\citenamefont{Stoof et~al.}(2008)\citenamefont{Stoof, Dickerscheid,
  and Gubbels}}]{Stoof:2008ho}
\bibinfo{author}{\bibfnamefont{H.~T.~C.} \bibnamefont{Stoof}},
  \bibinfo{author}{\bibfnamefont{D.~B.~M.} \bibnamefont{Dickerscheid}},
  \bibnamefont{and} \bibinfo{author}{\bibfnamefont{K.}~\bibnamefont{Gubbels}},
  \emph{\bibinfo{title}{{Ultracold Quantum Fields}}}, Theoretical and
  Mathematical Physics (\bibinfo{publisher}{Springer Netherlands},
  \bibinfo{address}{Dordrecht}, \bibinfo{year}{2008}).

\bibitem[{\citenamefont{Rab et~al.}(2008)\citenamefont{Rab, Cole, Parker,
  Greentree, Hollenberg, and Martin}}]{Rab:2008kp}
\bibinfo{author}{\bibfnamefont{M.}~\bibnamefont{Rab}},
  \bibinfo{author}{\bibfnamefont{J.~H.} \bibnamefont{Cole}},
  \bibinfo{author}{\bibfnamefont{N.~G.} \bibnamefont{Parker}},
  \bibinfo{author}{\bibfnamefont{A.~D.} \bibnamefont{Greentree}},
  \bibinfo{author}{\bibfnamefont{L.~C.~L.} \bibnamefont{Hollenberg}},
  \bibnamefont{and} \bibinfo{author}{\bibfnamefont{A.~M.}
  \bibnamefont{Martin}}, \bibinfo{journal}{Phys. Rev. A}
  \textbf{\bibinfo{volume}{77}}, \bibinfo{pages}{061602}
  (\bibinfo{year}{2008}).

\bibitem[{\citenamefont{Bradly et~al.}(2012)\citenamefont{Bradly, Rab,
  Greentree, and Martin}}]{Bradly:2012fc}
\bibinfo{author}{\bibfnamefont{C.~J.} \bibnamefont{Bradly}},
  \bibinfo{author}{\bibfnamefont{M.}~\bibnamefont{Rab}},
  \bibinfo{author}{\bibfnamefont{A.~D.} \bibnamefont{Greentree}},
  \bibnamefont{and} \bibinfo{author}{\bibfnamefont{A.~M.}
  \bibnamefont{Martin}}, \bibinfo{journal}{Phys. Rev. A}
  \textbf{\bibinfo{volume}{85}}, \bibinfo{pages}{053609}
  (\bibinfo{year}{2012}).

\bibitem[{\citenamefont{Muruganandam and Adhikari}(2009)}]{Muruganandam:2009dq}
\bibinfo{author}{\bibfnamefont{P.}~\bibnamefont{Muruganandam}}
  \bibnamefont{and} \bibinfo{author}{\bibfnamefont{S.~K.}
  \bibnamefont{Adhikari}}, \bibinfo{journal}{Comput. Phys. Commun.}
  \textbf{\bibinfo{volume}{180}}, \bibinfo{pages}{1888} (\bibinfo{year}{2009}).

\bibitem[{\citenamefont{Cerimele et~al.}(2000)\citenamefont{Cerimele, Chiofalo,
  Pistella, Succi, and Tosi}}]{Cerimele:2000hs}
\bibinfo{author}{\bibfnamefont{M.~M.} \bibnamefont{Cerimele}},
  \bibinfo{author}{\bibfnamefont{M.}~\bibnamefont{Chiofalo}},
  \bibinfo{author}{\bibfnamefont{F.}~\bibnamefont{Pistella}},
  \bibinfo{author}{\bibfnamefont{S.}~\bibnamefont{Succi}}, \bibnamefont{and}
  \bibinfo{author}{\bibfnamefont{M.}~\bibnamefont{Tosi}},
  \bibinfo{journal}{Phys. Rev. E} \textbf{\bibinfo{volume}{62}},
  \bibinfo{pages}{1382} (\bibinfo{year}{2000}).

\bibitem[{\citenamefont{Landauer}(1957)}]{Landauer:cw}
\bibinfo{author}{\bibfnamefont{R.}~\bibnamefont{Landauer}},
  \bibinfo{journal}{IBM J. Res. Dev.} \textbf{\bibinfo{volume}{1}},
  \bibinfo{pages}{223} (\bibinfo{year}{1957}).

\bibitem[{\citenamefont{Sakurai and Napolitano}(2011)}]{sakurai2011modern}
\bibinfo{author}{\bibfnamefont{J.~J.} \bibnamefont{Sakurai}} \bibnamefont{and}
  \bibinfo{author}{\bibfnamefont{J.}~\bibnamefont{Napolitano}},
  \emph{\bibinfo{title}{{Modern Quantum Mechanics}}}, 物理学经典教材
  (\bibinfo{publisher}{Addison-Wesley}, \bibinfo{year}{2011}).

\bibitem[{\citenamefont{Dekorsy et~al.}(1995)\citenamefont{Dekorsy, Ott, Kurz,
  and K{\"o}hler}}]{Dekorsy:1995ff}
\bibinfo{author}{\bibfnamefont{T.}~\bibnamefont{Dekorsy}},
  \bibinfo{author}{\bibfnamefont{R.}~\bibnamefont{Ott}},
  \bibinfo{author}{\bibfnamefont{H.}~\bibnamefont{Kurz}}, \bibnamefont{and}
  \bibinfo{author}{\bibfnamefont{K.}~\bibnamefont{K{\"o}hler}},
  \bibinfo{journal}{Phys. Rev. B} \textbf{\bibinfo{volume}{51}},
  \bibinfo{pages}{17275} (\bibinfo{year}{1995}).

\bibitem[{\citenamefont{Flayac et~al.}(2011)\citenamefont{Flayac, Solnyshkov,
  and Malpuech}}]{Flayac:2011ft}
\bibinfo{author}{\bibfnamefont{H.}~\bibnamefont{Flayac}},
  \bibinfo{author}{\bibfnamefont{D.~D.} \bibnamefont{Solnyshkov}},
  \bibnamefont{and} \bibinfo{author}{\bibfnamefont{G.}~\bibnamefont{Malpuech}},
  \bibinfo{journal}{Phys. Rev. B} \textbf{\bibinfo{volume}{83}},
  \bibinfo{pages}{045412} (\bibinfo{year}{2011}).

\bibitem[{\citenamefont{Greif et~al.}(2016)\citenamefont{Greif, Parsons,
  Mazurenko, and Chiu}}]{Greif:2015wi}
\bibinfo{author}{\bibfnamefont{D.}~\bibnamefont{Greif}},
  \bibinfo{author}{\bibfnamefont{M.~F.} \bibnamefont{Parsons}},
  \bibinfo{author}{\bibfnamefont{A.}~\bibnamefont{Mazurenko}},
  \bibnamefont{and} \bibinfo{author}{\bibfnamefont{C.~S.} \bibnamefont{Chiu}},
  \bibinfo{journal}{Science} \textbf{\bibinfo{volume}{351}},
  \bibinfo{pages}{953} (\bibinfo{year}{2016}).

\bibitem[{\citenamefont{Parsons et~al.}(2016)\citenamefont{Parsons, Mazurenko,
  Chiu, Ji, Greif, and Greiner}}]{Parsons:2016gr}
\bibinfo{author}{\bibfnamefont{M.~F.} \bibnamefont{Parsons}},
  \bibinfo{author}{\bibfnamefont{A.}~\bibnamefont{Mazurenko}},
  \bibinfo{author}{\bibfnamefont{C.~S.} \bibnamefont{Chiu}},
  \bibinfo{author}{\bibfnamefont{G.}~\bibnamefont{Ji}},
  \bibinfo{author}{\bibfnamefont{D.}~\bibnamefont{Greif}}, \bibnamefont{and}
  \bibinfo{author}{\bibfnamefont{M.}~\bibnamefont{Greiner}},
  \bibinfo{journal}{Science} \textbf{\bibinfo{volume}{353}},
  \bibinfo{pages}{1253} (\bibinfo{year}{2016}).

\bibitem[{\citenamefont{Cheuk et~al.}(2016)\citenamefont{Cheuk, Nichols,
  Lawrence, Okan, Zhang, and Zwierlein}}]{Cheuk:2016tp}
\bibinfo{author}{\bibfnamefont{L.~W.} \bibnamefont{Cheuk}},
  \bibinfo{author}{\bibfnamefont{M.~A.} \bibnamefont{Nichols}},
  \bibinfo{author}{\bibfnamefont{K.~R.} \bibnamefont{Lawrence}},
  \bibinfo{author}{\bibfnamefont{M.}~\bibnamefont{Okan}},
  \bibinfo{author}{\bibfnamefont{H.}~\bibnamefont{Zhang}}, \bibnamefont{and}
  \bibinfo{author}{\bibfnamefont{M.~W.} \bibnamefont{Zwierlein}},
  \bibinfo{journal}{Phys. Rev. Lett.} \textbf{\bibinfo{volume}{116}},
  \bibinfo{pages}{235301} (\bibinfo{year}{2016}).

\bibitem[{\citenamefont{Leder et~al.}(2016)\citenamefont{Leder, Grossert,
  Sitta, and Genske}}]{Leder:2016wy}
\bibinfo{author}{\bibfnamefont{M.}~\bibnamefont{Leder}},
  \bibinfo{author}{\bibfnamefont{C.}~\bibnamefont{Grossert}},
  \bibinfo{author}{\bibfnamefont{L.}~\bibnamefont{Sitta}}, \bibnamefont{and}
  \bibinfo{author}{\bibfnamefont{M.}~\bibnamefont{Genske}},
  \bibinfo{journal}{Nature (London)} \textbf{\bibinfo{volume}{7}},
  \bibinfo{pages}{13112} (\bibinfo{year}{2016}).

\bibitem[{\citenamefont{Miranda et~al.}(2015)\citenamefont{Miranda, Inoue,
  Okuyama, Nakamoto, and Kozuma}}]{Miranda:2014wi}
\bibinfo{author}{\bibfnamefont{M.}~\bibnamefont{Miranda}},
  \bibinfo{author}{\bibfnamefont{R.}~\bibnamefont{Inoue}},
  \bibinfo{author}{\bibfnamefont{Y.}~\bibnamefont{Okuyama}},
  \bibinfo{author}{\bibfnamefont{A.}~\bibnamefont{Nakamoto}}, \bibnamefont{and}
  \bibinfo{author}{\bibfnamefont{M.}~\bibnamefont{Kozuma}},
  \bibinfo{journal}{Phys. Rev. A} \textbf{\bibinfo{volume}{91}},
  \bibinfo{pages}{063414} (\bibinfo{year}{2015}).

\bibitem[{\citenamefont{Chien et~al.}(2012)\citenamefont{Chien, Zwolak, and
  Di~Ventra}}]{Chien:2012ft}
\bibinfo{author}{\bibfnamefont{C.-C.} \bibnamefont{Chien}},
  \bibinfo{author}{\bibfnamefont{M.}~\bibnamefont{Zwolak}}, \bibnamefont{and}
  \bibinfo{author}{\bibfnamefont{M.}~\bibnamefont{Di~Ventra}},
  \bibinfo{journal}{Phys. Rev. A} \textbf{\bibinfo{volume}{85}},
  \bibinfo{pages}{041601} (\bibinfo{year}{2012}).

\bibitem[{\citenamefont{Bushong et~al.}(2005)\citenamefont{Bushong, Sai, and
  Di~Ventra}}]{Bushong:2005kx}
\bibinfo{author}{\bibfnamefont{N.}~\bibnamefont{Bushong}},
  \bibinfo{author}{\bibfnamefont{N.}~\bibnamefont{Sai}}, \bibnamefont{and}
  \bibinfo{author}{\bibfnamefont{M.}~\bibnamefont{Di~Ventra}},
  \bibinfo{journal}{Nano Lett.} \textbf{\bibinfo{volume}{5}},
  \bibinfo{pages}{2569} (\bibinfo{year}{2005}).

\bibitem[{\citenamefont{McKay et~al.}(2008)\citenamefont{McKay, White,
  Pasienski, and DeMarco}}]{McKay:2008kt}
\bibinfo{author}{\bibfnamefont{D.}~\bibnamefont{McKay}},
  \bibinfo{author}{\bibfnamefont{M.}~\bibnamefont{White}},
  \bibinfo{author}{\bibfnamefont{M.}~\bibnamefont{Pasienski}},
  \bibnamefont{and} \bibinfo{author}{\bibfnamefont{B.}~\bibnamefont{DeMarco}},
  \bibinfo{journal}{Nature (London)} \textbf{\bibinfo{volume}{453}},
  \bibinfo{pages}{76} (\bibinfo{year}{2008}).

\bibitem[{\citenamefont{Abbate et~al.}(2017)\citenamefont{Abbate, Gori,
  Inguscio, Modugno, and D'Errico}}]{Abbate2017}
\bibinfo{author}{\bibfnamefont{S.~S.} \bibnamefont{Abbate}},
  \bibinfo{author}{\bibfnamefont{L.}~\bibnamefont{Gori}},
  \bibinfo{author}{\bibfnamefont{M.}~\bibnamefont{Inguscio}},
  \bibinfo{author}{\bibfnamefont{G.}~\bibnamefont{Modugno}}, \bibnamefont{and}
  \bibinfo{author}{\bibfnamefont{C.}~\bibnamefont{D'Errico}},
  \bibinfo{journal}{The European Physical Journal Special Topics}
  \textbf{\bibinfo{volume}{226}}, \bibinfo{pages}{2815} (\bibinfo{year}{2017}),
  ISSN \bibinfo{issn}{1951-6401}.

\bibitem[{\citenamefont{Jaksch and Zoller}(2005)}]{Jaksch:2005go}
\bibinfo{author}{\bibfnamefont{D.}~\bibnamefont{Jaksch}} \bibnamefont{and}
  \bibinfo{author}{\bibfnamefont{P.}~\bibnamefont{Zoller}},
  \bibinfo{journal}{Ann. Phys.} \textbf{\bibinfo{volume}{315}},
  \bibinfo{pages}{52} (\bibinfo{year}{2005}).

\bibitem[{\citenamefont{Bloch et~al.}(2008)\citenamefont{Bloch, Dalibard, and
  Zwerger}}]{Bloch:2008gl}
\bibinfo{author}{\bibfnamefont{I.}~\bibnamefont{Bloch}},
  \bibinfo{author}{\bibfnamefont{J.}~\bibnamefont{Dalibard}}, \bibnamefont{and}
  \bibinfo{author}{\bibfnamefont{W.}~\bibnamefont{Zwerger}},
  \bibinfo{journal}{Rev. Mod. Phys.} \textbf{\bibinfo{volume}{80}},
  \bibinfo{pages}{885} (\bibinfo{year}{2008}).

\bibitem[{\citenamefont{Morsch et~al.}(2002)\citenamefont{Morsch, Cristiani,
  M{\"u}ller, Ciampini, and Arimondo}}]{Morsch:2002dk}
\bibinfo{author}{\bibfnamefont{O.}~\bibnamefont{Morsch}},
  \bibinfo{author}{\bibfnamefont{M.}~\bibnamefont{Cristiani}},
  \bibinfo{author}{\bibfnamefont{J.~H.} \bibnamefont{M{\"u}ller}},
  \bibinfo{author}{\bibfnamefont{D.}~\bibnamefont{Ciampini}}, \bibnamefont{and}
  \bibinfo{author}{\bibfnamefont{E.}~\bibnamefont{Arimondo}},
  \bibinfo{journal}{Phys. Rev. A} \textbf{\bibinfo{volume}{66}},
  \bibinfo{pages}{021601} (\bibinfo{year}{2002}).

\bibitem[{\citenamefont{Esslinger}(2010)}]{Esslinger:2010ex}
\bibinfo{author}{\bibfnamefont{T.}~\bibnamefont{Esslinger}},
  \bibinfo{journal}{Annu. Rev. Condens. Matter Phys.}
  \textbf{\bibinfo{volume}{1}}, \bibinfo{pages}{129} (\bibinfo{year}{2010}).

\bibitem[{\citenamefont{Chin et~al.}(2010)\citenamefont{Chin, Grimm, Julienne,
  and Tiesinga}}]{Chin:2010kl}
\bibinfo{author}{\bibfnamefont{C.}~\bibnamefont{Chin}},
  \bibinfo{author}{\bibfnamefont{R.}~\bibnamefont{Grimm}},
  \bibinfo{author}{\bibfnamefont{P.~S.} \bibnamefont{Julienne}},
  \bibnamefont{and} \bibinfo{author}{\bibfnamefont{E.}~\bibnamefont{Tiesinga}},
  \bibinfo{journal}{Rev. Mod. Phys.} \textbf{\bibinfo{volume}{82}},
  \bibinfo{pages}{1225} (\bibinfo{year}{2010}).

\bibitem[{\citenamefont{Schunck et~al.}(2005)\citenamefont{Schunck, Zwierlein,
  Stan, Raupach, Ketterle, Simoni, Tiesinga, Williams, and
  Julienne}}]{Schunck:2005cf}
\bibinfo{author}{\bibfnamefont{C.~H.} \bibnamefont{Schunck}},
  \bibinfo{author}{\bibfnamefont{M.~W.} \bibnamefont{Zwierlein}},
  \bibinfo{author}{\bibfnamefont{C.~A.} \bibnamefont{Stan}},
  \bibinfo{author}{\bibfnamefont{S.~M.~F.} \bibnamefont{Raupach}},
  \bibinfo{author}{\bibfnamefont{W.}~\bibnamefont{Ketterle}},
  \bibinfo{author}{\bibfnamefont{A.}~\bibnamefont{Simoni}},
  \bibinfo{author}{\bibfnamefont{E.}~\bibnamefont{Tiesinga}},
  \bibinfo{author}{\bibfnamefont{C.~J.} \bibnamefont{Williams}},
  \bibnamefont{and} \bibinfo{author}{\bibfnamefont{P.~S.}
  \bibnamefont{Julienne}}, \bibinfo{journal}{Phys. Rev. A}
  \textbf{\bibinfo{volume}{71}}, \bibinfo{pages}{045601}
  (\bibinfo{year}{2005}).

\bibitem[{\citenamefont{Clark et~al.}(2015)\citenamefont{Clark, Ha, Xu, and
  Chin}}]{Clark:2015uj}
\bibinfo{author}{\bibfnamefont{L.~W.} \bibnamefont{Clark}},
  \bibinfo{author}{\bibfnamefont{L.~C.} \bibnamefont{Ha}},
  \bibinfo{author}{\bibfnamefont{C.~Y.} \bibnamefont{Xu}}, \bibnamefont{and}
  \bibinfo{author}{\bibfnamefont{C.}~\bibnamefont{Chin}},
  \bibinfo{journal}{Phys. Rev. Lett.} \textbf{\bibinfo{volume}{115}},
  \bibinfo{pages}{155301} (\bibinfo{year}{2015}).

\bibitem[{\citenamefont{Fu et~al.}(2013)\citenamefont{Fu, Wang, Huang, Meng,
  Hu, and Zhang}}]{Fu:2013im}
\bibinfo{author}{\bibfnamefont{Z.}~\bibnamefont{Fu}},
  \bibinfo{author}{\bibfnamefont{P.}~\bibnamefont{Wang}},
  \bibinfo{author}{\bibfnamefont{L.}~\bibnamefont{Huang}},
  \bibinfo{author}{\bibfnamefont{Z.}~\bibnamefont{Meng}},
  \bibinfo{author}{\bibfnamefont{H.}~\bibnamefont{Hu}}, \bibnamefont{and}
  \bibinfo{author}{\bibfnamefont{J.}~\bibnamefont{Zhang}},
  \bibinfo{journal}{Phys. Rev. A} \textbf{\bibinfo{volume}{88}},
  \bibinfo{pages}{041601} (\bibinfo{year}{2013}).

\bibitem[{\citenamefont{Yamazaki et~al.}(2010)\citenamefont{Yamazaki, Taie,
  Sugawa, and Takahashi}}]{Yamazaki:2010en}
\bibinfo{author}{\bibfnamefont{R.}~\bibnamefont{Yamazaki}},
  \bibinfo{author}{\bibfnamefont{S.}~\bibnamefont{Taie}},
  \bibinfo{author}{\bibfnamefont{S.}~\bibnamefont{Sugawa}}, \bibnamefont{and}
  \bibinfo{author}{\bibfnamefont{Y.}~\bibnamefont{Takahashi}},
  \bibinfo{journal}{Phys. Rev. Lett.} \textbf{\bibinfo{volume}{105}},
  \bibinfo{pages}{050405} (\bibinfo{year}{2010}).

\bibitem[{\citenamefont{Fatemi et~al.}(2000)\citenamefont{Fatemi, Jones, and
  Lett}}]{Fatemi:2000iy}
\bibinfo{author}{\bibfnamefont{F.~K.} \bibnamefont{Fatemi}},
  \bibinfo{author}{\bibfnamefont{K.~M.} \bibnamefont{Jones}}, \bibnamefont{and}
  \bibinfo{author}{\bibfnamefont{P.~D.} \bibnamefont{Lett}},
  \bibinfo{journal}{Phys. Rev. Lett.} \textbf{\bibinfo{volume}{85}},
  \bibinfo{pages}{4462} (\bibinfo{year}{2000}).

\bibitem[{\citenamefont{Enomoto et~al.}(2008)\citenamefont{Enomoto, Kasa,
  Kitagawa, and Takahashi}}]{Enomoto:2008bx}
\bibinfo{author}{\bibfnamefont{K.}~\bibnamefont{Enomoto}},
  \bibinfo{author}{\bibfnamefont{K.}~\bibnamefont{Kasa}},
  \bibinfo{author}{\bibfnamefont{M.}~\bibnamefont{Kitagawa}}, \bibnamefont{and}
  \bibinfo{author}{\bibfnamefont{Y.}~\bibnamefont{Takahashi}},
  \bibinfo{journal}{Phys. Rev. Lett.} \textbf{\bibinfo{volume}{101}},
  \bibinfo{pages}{203201} (\bibinfo{year}{2008}).

\bibitem[{\citenamefont{Fukuhara et~al.}(2009)\citenamefont{Fukuhara, Sugawa,
  Takasu, and Takahashi}}]{Fukuhara:2009bv}
\bibinfo{author}{\bibfnamefont{T.}~\bibnamefont{Fukuhara}},
  \bibinfo{author}{\bibfnamefont{S.}~\bibnamefont{Sugawa}},
  \bibinfo{author}{\bibfnamefont{Y.}~\bibnamefont{Takasu}}, \bibnamefont{and}
  \bibinfo{author}{\bibfnamefont{Y.}~\bibnamefont{Takahashi}},
  \bibinfo{journal}{Phys. Rev. A} \textbf{\bibinfo{volume}{79}},
  \bibinfo{pages}{021601} (\bibinfo{year}{2009}).

\bibitem[{\citenamefont{Wu and Thomas}(2012)}]{Wu:2012jp}
\bibinfo{author}{\bibfnamefont{H.}~\bibnamefont{Wu}} \bibnamefont{and}
  \bibinfo{author}{\bibfnamefont{J.~E.} \bibnamefont{Thomas}},
  \bibinfo{journal}{Phys. Rev. Lett.} \textbf{\bibinfo{volume}{108}},
  \bibinfo{pages}{010401} (\bibinfo{year}{2012}).

\bibitem[{\citenamefont{Stock and Deutsch}(2006)}]{Stock:2006gw}
\bibinfo{author}{\bibfnamefont{R.}~\bibnamefont{Stock}} \bibnamefont{and}
  \bibinfo{author}{\bibfnamefont{I.~H.} \bibnamefont{Deutsch}},
  \bibinfo{journal}{Phys. Rev. A} \textbf{\bibinfo{volume}{73}},
  \bibinfo{pages}{032701} (\bibinfo{year}{2006}).

\bibitem[{\citenamefont{Moritz et~al.}(2005)\citenamefont{Moritz, St{\"o}ferle,
  G{\"u}nter, K{\"o}hl, and Esslinger}}]{Moritz:2005bf}
\bibinfo{author}{\bibfnamefont{H.}~\bibnamefont{Moritz}},
  \bibinfo{author}{\bibfnamefont{T.}~\bibnamefont{St{\"o}ferle}},
  \bibinfo{author}{\bibfnamefont{K.}~\bibnamefont{G{\"u}nter}},
  \bibinfo{author}{\bibfnamefont{M.}~\bibnamefont{K{\"o}hl}}, \bibnamefont{and}
  \bibinfo{author}{\bibfnamefont{T.}~\bibnamefont{Esslinger}},
  \bibinfo{journal}{Phys. Rev. Lett.} \textbf{\bibinfo{volume}{94}},
  \bibinfo{pages}{210401} (\bibinfo{year}{2005}).

\bibitem[{\citenamefont{Sekh}(2012)}]{Sekh:2012fn}
\bibinfo{author}{\bibfnamefont{G.~A.} \bibnamefont{Sekh}},
  \bibinfo{journal}{Physics Letters A} \textbf{\bibinfo{volume}{376}},
  \bibinfo{pages}{1740} (\bibinfo{year}{2012}).

\bibitem[{\citenamefont{Tiesinga et~al.}(2000)\citenamefont{Tiesinga, Williams,
  Mies, and Julienne}}]{Tiesinga:2000fm}
\bibinfo{author}{\bibfnamefont{E.}~\bibnamefont{Tiesinga}},
  \bibinfo{author}{\bibfnamefont{C.~J.} \bibnamefont{Williams}},
  \bibinfo{author}{\bibfnamefont{F.~H.} \bibnamefont{Mies}}, \bibnamefont{and}
  \bibinfo{author}{\bibfnamefont{P.~S.} \bibnamefont{Julienne}},
  \bibinfo{journal}{Phys. Rev. A} \textbf{\bibinfo{volume}{61}},
  \bibinfo{pages}{063416} (\bibinfo{year}{2000}).

\bibitem[{\citenamefont{Huber and R{\"u}egg}(2009)}]{Huber:2009bu}
\bibinfo{author}{\bibfnamefont{S.~D.} \bibnamefont{Huber}} \bibnamefont{and}
  \bibinfo{author}{\bibfnamefont{A.}~\bibnamefont{R{\"u}egg}},
  \bibinfo{journal}{Phys. Rev. Lett.} \textbf{\bibinfo{volume}{102}},
  \bibinfo{pages}{065301} (\bibinfo{year}{2009}).

\bibitem[{\citenamefont{Strohmaier et~al.}(2010)\citenamefont{Strohmaier,
  Greif, J{\"o}rdens, Tarruell, Moritz, Esslinger, Sensarma, Pekker, Altman,
  and Demler}}]{Strohmaier:2010fr}
\bibinfo{author}{\bibfnamefont{N.}~\bibnamefont{Strohmaier}},
  \bibinfo{author}{\bibfnamefont{D.}~\bibnamefont{Greif}},
  \bibinfo{author}{\bibfnamefont{R.}~\bibnamefont{J{\"o}rdens}},
  \bibinfo{author}{\bibfnamefont{L.}~\bibnamefont{Tarruell}},
  \bibinfo{author}{\bibfnamefont{H.}~\bibnamefont{Moritz}},
  \bibinfo{author}{\bibfnamefont{T.}~\bibnamefont{Esslinger}},
  \bibinfo{author}{\bibfnamefont{R.}~\bibnamefont{Sensarma}},
  \bibinfo{author}{\bibfnamefont{D.}~\bibnamefont{Pekker}},
  \bibinfo{author}{\bibfnamefont{E.}~\bibnamefont{Altman}}, \bibnamefont{and}
  \bibinfo{author}{\bibfnamefont{E.}~\bibnamefont{Demler}},
  \bibinfo{journal}{Phys. Rev. Lett.} \textbf{\bibinfo{volume}{104}},
  \bibinfo{pages}{080401} (\bibinfo{year}{2010}).

\bibitem[{\citenamefont{Covey et~al.}(2016)\citenamefont{Covey, Moses,
  G{\"a}rttner, Safavi-Naini, Miecnikowski, Fu, Schachenmayer, Julienne, Rey,
  Jin et~al.}}]{Covey:2016bc}
\bibinfo{author}{\bibfnamefont{J.~P.} \bibnamefont{Covey}},
  \bibinfo{author}{\bibfnamefont{S.~A.} \bibnamefont{Moses}},
  \bibinfo{author}{\bibfnamefont{M.}~\bibnamefont{G{\"a}rttner}},
  \bibinfo{author}{\bibfnamefont{A.}~\bibnamefont{Safavi-Naini}},
  \bibinfo{author}{\bibfnamefont{M.~T.} \bibnamefont{Miecnikowski}},
  \bibinfo{author}{\bibfnamefont{Z.}~\bibnamefont{Fu}},
  \bibinfo{author}{\bibfnamefont{J.}~\bibnamefont{Schachenmayer}},
  \bibinfo{author}{\bibfnamefont{P.~S.} \bibnamefont{Julienne}},
  \bibinfo{author}{\bibfnamefont{A.~M.} \bibnamefont{Rey}},
  \bibinfo{author}{\bibfnamefont{D.~S.} \bibnamefont{Jin}},
  \bibnamefont{et~al.}, \bibinfo{journal}{Nat. Commun.}
  \textbf{\bibinfo{volume}{7}}, \bibinfo{pages}{11279}
  (\bibinfo{year}{2016}).

\bibitem[{\citenamefont{Rausch and Potthoff}(2017)}]{Rausch:2017ck}
\bibinfo{author}{\bibfnamefont{R.}~\bibnamefont{Rausch}} \bibnamefont{and}
  \bibinfo{author}{\bibfnamefont{M.}~\bibnamefont{Potthoff}},
  \bibinfo{journal}{Phys. Rev. B} \textbf{\bibinfo{volume}{95}},
  \bibinfo{pages}{045152} (\bibinfo{year}{2017}).

\bibitem[{\citenamefont{Ryu et~al.}(2013)\citenamefont{Ryu, Blackburn, Blinova,
  and Boshier}}]{PhysRevLett.111.205301}
\bibinfo{author}{\bibfnamefont{C.}~\bibnamefont{Ryu}},
  \bibinfo{author}{\bibfnamefont{P.~W.} \bibnamefont{Blackburn}},
  \bibinfo{author}{\bibfnamefont{A.~A.} \bibnamefont{Blinova}},
  \bibnamefont{and} \bibinfo{author}{\bibfnamefont{M.~G.}
  \bibnamefont{Boshier}}, \bibinfo{journal}{Phys. Rev. Lett.}
  \textbf{\bibinfo{volume}{111}}, \bibinfo{pages}{205301}
  (\bibinfo{year}{2013}).

\bibitem[{\citenamefont{Kao et~al.}(2015)\citenamefont{Kao, Hsieh, and
  Chen}}]{Kao:2015gb}
\bibinfo{author}{\bibfnamefont{Y.~J.} \bibnamefont{Kao}},
  \bibinfo{author}{\bibfnamefont{Y.~D.} \bibnamefont{Hsieh}}, \bibnamefont{and}
  \bibinfo{author}{\bibfnamefont{P.}~\bibnamefont{Chen}},
  \bibinfo{journal}{Journal of Physics: Conference Series}
  \textbf{\bibinfo{volume}{640}}, \bibinfo{pages}{012040}
  (\bibinfo{year}{2015}).

\end{thebibliography}

\end{document}